\documentclass[aps,prc,superscriptaddress]{revtex4}

\newcommand{\be}{\begin{eqnarray}}
\newcommand{\ee}{\end{eqnarray}}
\newcommand{\gs}{g_{\rm s}}

\newcommand{\p}{{\bf p}}
\newcommand{\F}{{\bf F}}
\newcommand{\h}{{\bf h}}

\usepackage{epsfig,amsmath}

\begin{document}
\pagestyle{empty}
\title{High Momentum Dilepton Production from Jets in a Quark Gluon Plasma}
\author{Simon Turbide}
\author{Charles Gale}
\affiliation
{Department of Physics, McGill University, 3600 University Street,
    Montreal, Canada H3A 2T8}

\author{Dinesh K. Srivastava}
\affiliation{Variable Energy Cyclotron Centre, 1/AF Bidhan Nagar, 
Kolkata 700 064, India}

\author{Rainer J. Fries}
\affiliation{School of Physics and Astronomy, University of Minnesota, 
Minneapolis, MN 55455}

\date{\today}

\begin{abstract}
We discuss the emission of high momentum lepton pairs in central
Au+Au collisions at RHIC ($\sqrt{s_{NN}}=$200 GeV) and Pb+Pb collisions at the
LHC ($\sqrt{s_{NN}}=$5500 GeV). Yields of dileptons produced through 
interactions of jets with thermal partons have been calculated, with 
next-to-leading order corrections through hard thermal loop (HTL)
resummation. They are compared to thermal dilepton emission and the 
Drell-Yan process. A complete leading order treatment of jet energy loss has 
been included.  Jet-plasma interactions are found to dominate over 
thermal dilepton emission for all values of the invariant mass $M$. 
Drell-Yan is the dominant source of high momentum lepton pairs for 
$M >$ 3 GeV at RHIC, after the background from heavy quark decays is 
subtracted. At LHC, the range $M<7$ GeV is dominated by jet-plasma 
interactions. Effects from jet energy loss on jet-plasma interactions turn 
out to be weak, but non-negligible, reducing the yield of low-mass dileptons 
by about $30\%$.

\end{abstract}

 \maketitle


\pagestyle{plain}

\section{Introduction}

Finding experimental evidence for the existence of a Quark-Gluon Plasma 
(QGP) is one of the main reasons for conducting relativistic heavy ion 
collisions. In this context, real and virtual photons can be used to probe the high 
temperature and density phase in these collisions: due to their large 
mean free path, electromagnetic radiation suffers only little final 
state interactions~\cite{Feinb:76,Shuryak:1978ij}.  More recently, photons emitted from jets 
interacting with the medium have been suggested as a further signature 
\cite{prlphoton}. A quantitative characterization of the different sources 
contributing to the photon yield in high energy heavy ion collisions 
can be found in \cite{simon2}. In this work, we concentrate on virtual 
photons decaying into lepton pairs.

Several sources for lepton pairs compete and have to be considered: dileptons from 
the Drell-Yan process~\cite{drell-yan}, thermal dileptons~\cite{kkmm} 
both from the QGP~\cite{kms,altherr} and from the subsequent hadronic 
phase~\cite{gale_lichard,rapp_wam}, dileptons from the absorption of jets by 
the plasma \cite{prcdilep} and dileptons from bremsstrahlung of jets. 
Besides the thermal emission, dileptons from the latter two sources 
carry information about the medium. Comparing with the baseline 
$p+p$ process, we note that absorption of jets exclusively happens in 
nucleus-nucleus collisions in the presence of a medium. There is a also a 
significant change in bremsstrahlung emission expected in a medium (in A+A)
compared to the case of vacuum (in $p+p$).

The main background for dileptons are correlated charm decays 
($D\bar{D}\rightarrow e^+e^-X$) at intermediate mass~\cite{Shor:1989yt}, while one has to cope
with Dalitz decays of light mesons 
($\pi^0\rightarrow\gamma e^+e^-,\omega \rightarrow \pi^0e^+e^-$) at low mass 
$M\leq $ 1 GeV.  In principle, there is also another source of dileptons 
corresponding to pre-equilibrium emission.  It is difficult to assess both 
theoretically and experimentally, nevertheless, such calculations have been
attempted~\cite{geiger}.

In Ref.~\cite{prcdilep} the dilepton yield from the passage of jets through a 
plasma has been evaluated as a function of invariant mass at leading order.
The results show that this process dominates over thermally induced reactions 
at high invariant mass $M$.  However, energy loss of jets in the medium
has not been included in this calculation. Large energy loss of jets was 
discovered as one of the most exciting results from the Relativistic 
Heavy Ion Collider (RHIC) and has been observed in the suppression of
high $p_T$ hadron spectra~\cite{phenix2,star} and through the 
disappearance of back-to-back correlations of
high $p_T$ hadrons~\cite{star2}.

In this paper we re-examine the leading order dilepton production from 
jets~\cite{note1} and we explore the effect of energy loss.  We assume here 
that jets lose their energy by induced gluon bremsstrahlung only~\cite{Gyu_plum,wgp}. 
We use the approach developed by Arnold, Moore, and Yaffe (AMY)~\cite{AMY}, 
which correctly treats the Landau-Pomeranchuk-Migdal (LPM) effect 
(up to $O(g_s)$ corrections)~\cite{migdal}. This model has been used 
successfully to reproduce the measured nuclear modification factor $R_{AA}$ of
neutral pions, the ratio of all photons over background photons and the direct
photon yield at RHIC \cite{simon2}.  
Jets will be defined by all partons produced initially with transverse
momentum $p_T^{\rm jet}\gg 1$ GeV. The total dilepton production could be
influenced by the choice of the cutoff $p_T^{\rm jet}$.  
As we will discuss below, in order to avoid such sensitivity, we limit 
our study to high momentum dileptons.  

In perturbation theory at high temperature, it is important to distinguish
between hard momenta, on the order of the temperature $T$, and soft momenta,
on the order of $g_sT$, where the QCD coupling constant is assumed to be small,
$g_s\ll 1$.  Only at the end of the calculations, will the results be
extrapolated to more realistic value of the coupling constant. When a 
line entering a vertex is soft, there are an infinite number of diagrams 
with loop corrections contributing to the same order in the coupling constant 
as the tree amplitude. These corrections can be treated with the hard thermal 
loop (HTL) resummation technique developed in Ref.~\cite{htl}. 
This technique has been used thereafter in 
Refs.~\cite{braaten_dil,Wong:1991be,thoma} to show that the production rate 
of low mass dileptons ($M < T$) could differ from the Born-term 
quark-antiquark annihilation by orders of magnitude.  In this work, we apply this technique to go beyond the leading order jet-medium interaction.

The dilepton production rate in finite-temperature field theory is calculated 
in Sec.~\ref{sec:ii} and the physical processes underlying each contribution are
discussed in Sec.~\ref{sec:iii} in the framework of relativistic kinetic 
theory.  In Sec.~\ref{sec:iv}, we include these dilepton production rates 
in a model for a longitudinally expanding QGP fireball. In Sec.~\ref{sec:v} 
we calculate the yield from the Drell-Yan process and correlated leptons 
from heavy quark decays. The numerical results are presented in 
Sec.~\ref{sec:vi}. Finally, Sec.~\ref{sec:vii} contains our summary and 
conclusions.

\section{Dilepton Production Rate from Finite-Temperature Field Theory}
\label{sec:ii}

Within the thermal field theory framework, the production rate of a 
lepton pair with momenta $p_+$ and $p_-$ is given by~\cite{vmd91}
\be
  E_+E_-\frac{dR^{e^+e^-}}{d^3p_+d^3p_-} = \frac{2 e^2}{(2\pi)^6(p_++p_-)^4}
  (p_+^{\mu}p_-^{\nu}+p_-^{\mu}p_+^{\nu}-g^{\mu\nu}p_+\cdot p_-)
  \frac{{\rm Im}\Pi^{R}_{\mu\nu}}{e^{(E_++E_-)/T}-1}
\ee
where $\Pi^{R}_{\mu\nu}$ is the finite-temperature retarded photon self-energy. After integration 
over the angular distribution of the lepton pair, we get
\be
  \frac{dR^{e^+e^-}}{d^4p} = 2E\frac{dR^{e^+e^-}}{dM^2d^3p}
  = \frac{2\alpha}{3\pi M^2}E\frac{dR^{\gamma^*}}{d^3p}
  = \frac{\alpha}{12\pi^4 M^2}\frac{{\rm Im}
  \Pi^{R \mu}_{ \mu}}{1-e^{E/T}}
\label{rate}
\ee
where $\vec{p}=\vec{p_+}+\vec{p_-}, E$ and $M$ are the momentum, energy and 
invariant mass of the virtual photon. 

\begin{figure}
\centerline{\epsfig{file=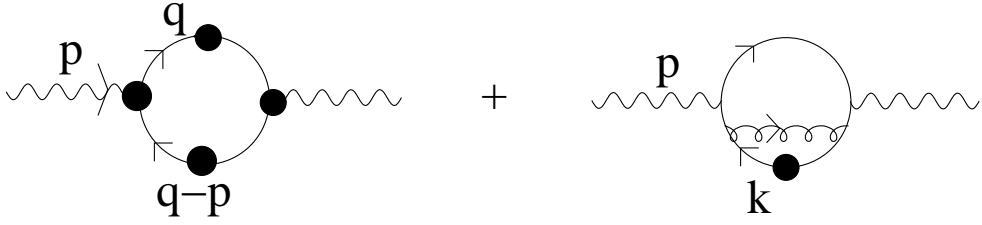,width=13cm}}
\caption{ Resummed photon self-energy diagrams.}
\label{pi_gamma}
\end{figure}

\begin{figure}
\centerline{\epsfig{file=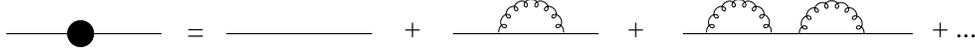,width=13cm}}
\caption{ Resummed quark propagator.}
\label{S_resum}
\end{figure}

In the HTL formalism, the leading order resummed photon self-energy 
diagram is shown in the left hand side of Fig.~\ref{pi_gamma}. The heavy 
dots indicate a resummed propagator or vertex. The second diagram in 
Fig.~\ref{pi_gamma}, coming from the effective two-photon-two quark vertex, 
is needed in order to fulfill the Ward identity
\be
p^{\mu}\Pi^{R}_{\mu\nu}(p)=0.
\ee
Using power counting~\cite{htl,lebellac}, the HTL resummation gives 
corrections of order $g_s^2T^2/|\vec{p}|^2$ to each bare 
vertex of the first diagram of Fig.~\ref{pi_gamma}, while the bare 
propagators receive $g_s^2T^2/|\vec{q}-\vec{p}|^2$ 
and $g_s^2T^2/|\vec{q}|^2$ corrections.
A resummed propagator consists of an infinite number of gluon correction, as
shown in Fig.~\ref{S_resum}. 

\begin{figure}
\centerline{\epsfig{file=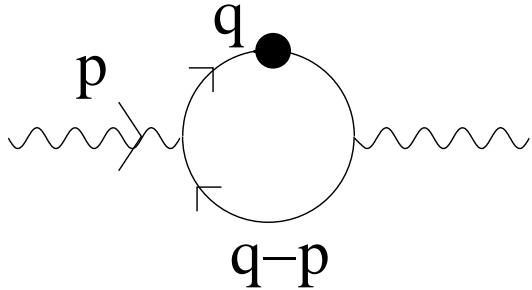,width=7cm}}
\caption{ Effective photon self-energy diagram for hard external momentum $p$.}
\label{pi_gamma_pt}
\end{figure}

In this work, we are interested in the  
high momentum dilepton limit ($|\vec{p}| \gg T$), so that vertex corrections 
can be neglected, and at least one propagator is guaranteed to be hard. 
We assume that this is the propagator associated with momentum $q-p$. 
Then the propagator associated with $q$ can be soft or hard, implying that 
is has to be resummed. Also, for the high $|\vec{p}|$ limit, the second 
diagram of Fig.~\ref{pi_gamma} gives only $g_s^2T^2/|\vec{p}|^2$ corrections 
to the first diagram.  The resulting diagram that has to be evaluated is
shown in Fig.~\ref{pi_gamma_pt} and its contribution to the self-energy is 
\be
\label{pi_1}
  \Pi^{\mu}_{\mu}(p) = 3e^2\sum_f\left(\frac{e_f}{e}\right)^2T\sum_n\int 
  \frac{d^3q}{(2\pi)^3}\mbox{Tr}\left[\gamma^{\mu}S_D(q)\gamma_{\mu}S(q-p) 
  \right]
\ee
with the sum running over the Matsubara frequencies. The diagram in Fig.~\ref{pi_gamma_pt} includes 
the leading order effect, and some next-to leading order corrections in $g_s$.
In order to have a complete next-to leading order production rate in the region $|\overrightarrow{p}|\gg T$, contributions like bremsstrahlung need to be included.  This will be discussed in the last section.

The dressed fermion propagator, in Euclidean space 
($\not\!q=-iq_0\gamma^0+\vec{q}\cdot\hat{\gamma}$), is given
  by~\cite{kapusta_book} 
\be
S_D(q)=\frac{1}{\not\!q+\Sigma}=\frac{\gamma^0-\hat{\gamma}\cdot \hat{\bf q}}{2D_+(q)}+\frac{\gamma^0+\hat{\gamma}\cdot \hat{\bf q}}{2D_-(q)}
\ee
where
\be
\label{D_def}
D_{\pm}(q)=-iq_0\pm|\overrightarrow{q}|+A\pm B
\ee
The terms $A$ and $B$ describe the quark self-energy  
\be
\Sigma=A\gamma^0+B\hat{\gamma}\cdot \hat{\bf q}.
\ee

In the HTL approximation, one obtains~\cite{lebellac}
\be
\label{A_B_def}
  A = \frac{m_F^2}{|\vec{q}|}Q_0\left(\frac{iq_0}{|\vec{q}|}\right),
  \quad B = -\frac{m_F^2}{|\vec{q}|}Q_1\left(\frac{iq_0}{|\vec{q}|}\right)
\ee
where $m_F=g_s T/\sqrt{6}$ is the effective quark mass induced by the thermal 
medium and the $Q_n$ are Legendre functions of the second kind. 
The bare propagator $S(q)$ is the same as $S_D(q)$, with $D_{\pm}$ 
replaced by
\be
  d_{\pm}(q)=-iq_0\pm|\vec{q}|
\ee
Using those expressions for the quark propagators and carrying out the trace, 
Eq.~\ref{pi_1} leads to
\be
  \Pi^{\mu}_{\mu}(p) &=& 6e^2\sum_f\left(\frac{e_f}{e}\right)^2 T 
  \sum_n\int \frac{d^3q}{(2\pi)^3} 
  \nonumber \\ && \times \Big[\frac{1}{D_+(q)}
  \left(\frac{1-\hat{\bf q}\cdot\hat{\bf k}}{d_+(k)}+\frac{1+\hat{\bf q}
  \cdot\hat{\bf k}}{d_-(k)}\right) 
   +\frac{1}{D_-(q)}\left(\frac{1+\hat{\bf q}\cdot\hat{\bf k}}{d_+(k)}
   +\frac{1-\hat{\bf q}\cdot\hat{\bf k}}{d_-(k)}\right) \Big]
\label{pi_2}
\ee
where we have defined $k=q-p$. 

At this point it is convenient to introduce 
the spectral representation of the effective quark propagator defined by
\be
\label{rho}
  \rho_{\pm}(\omega,\left|\vec{q}\right|)&=&-2\mbox{Im}\,\frac{1}{D_{\pm}(iq_0,
  \left|\vec{q}\right|)},\nonumber \\ 
  \frac{1}{D_{\pm}(iq_0,\left|\vec{q}\right|)}&=&
  \int_{-\infty}^{\infty}\frac{d\omega}{2\pi}
  \frac{\rho_{\pm}(\omega,\left|\vec{q}\right|)}{iq_0-\omega+i\epsilon}.
\ee
This implies
\be
\label{rho_2}
  \rho_{\pm}(\omega,\left|\vec{q}\right|) = &-&
  2\pi\frac{\omega^2-\left|\vec{q}\right|^2}{2m_F^2}\left[\delta
  \left(\omega-\omega_{\pm}(\left|\vec{q}\right|)\right)+\delta\left(\omega+\omega_{\mp}
  (\left|\vec{q}\right|)\right)\right] 
  -2\pi\beta_{\pm}\left(\omega,\left|\vec{q}\right|\right)\Theta
  \left(\left|\vec{q}\right|^2-\omega^2\right)
\ee
with 
\be
  \beta_{\pm}(\omega,\left|\vec{q}\right|) =
  \frac{\frac{m_F^2}{2\left|\vec{q}\right|}
  (1\mp \omega/\left|\vec{q}\right|)}{\left[-\omega\pm
  \left|\vec{q}\right|+\frac{m_F^2}{\left|\vec{q}\right|}
  \left(\pm1+\frac{1}{2}\ln
  \left|\frac{\omega+\left|\vec{q}\right|}{\omega-\left|\vec{q}\right|}  
  \right| \right)(1\mp \omega/\left|\vec{q}\right|) \right]^2+
  \left(\frac{\pi m_F^2}{2\left|\vec{q}\right|}(1\mp \omega/\left|\vec{q}\right|)  
  \right)^2}\, .
\ee
Here $\omega_{\pm}=\omega_{\pm}(\left|\vec{q}\right|)$ correspond to the 
poles of the effective quark propagator.  They are solution of 
$D_{\pm}(\omega,\left|\vec{q}\right|)=0$.  The solution $\omega=\omega_+$ 
represents an ordinary quark with an effective thermal mass $\sqrt{2}m_F$ 
and a positive helicity over chirality ratio, $\chi=1$~\cite{braaten_dil}.  
The solution $\omega=\omega_-$ represents a particle having negative 
helicity over chirality ratio, $\chi=-1$.  This collective mode, called 
plasmino, has no analog at zero temperature. Following the 
notation of Ref.~\cite{braaten_dil}, we denote the ordinary modes by $q_+$ 
and the plasminos by $q_-$.  The spectral density of the bare propagator is simply
\be 
  r_{\pm}(\omega',|\vec{k}|)&=&-2\mbox{Im}\,
  \frac{1}{d_{\pm}(ik_0,|\vec{k}|)}=-2\pi\delta(\omega'
  \mp |\vec{k}|)
\ee
Using the spectral density representation has the advantage that one can
carry out the sum over Matsubara frequencies with help of the elegant 
identity~\cite{braaten_dil},
\be
  \mbox{Im}\,T\sum_n F_1(iq_0)F_2(iq_0-ip_0) &=&
  \pi\left(1-e^{E/T}\right)\int_{-\infty}^{\infty}
  \frac{d\omega}{2\pi}\int_{-\infty}^{\infty}\frac{d\omega'}{2\pi}\nonumber
  \\ &&\times\rho_1(\omega)\rho_2(-\omega')\delta(E-\omega-\omega') 
  f_\mathrm{FD}(\omega)f_\mathrm{FD}(\omega')
\ee
where $f_\mathrm{FD}$ is the Fermi-Dirac phase space distribution function and 
$\rho_i$ is the spectral density associated with $F_i$. We use the analytical 
continuation $ip_0\rightarrow E+i\epsilon$ with the dilepton energy $E$.  

Putting all information together, we obtain
\be
  {\rm Im}\,\Pi^{R   \mu}_{ \mu} 
  &=&-6e^2 \pi\sum_f\left(\frac{e_f}{e}\right)^2 
  \left(1-e^{E/T}\right)\int\frac{d^3q}{(2\pi)^3}\int_{-\infty}^{\infty}
  \frac{d\omega}{2\pi}f_\mathrm{FD}(\omega)\nonumber \\
  && \times \Big[\rho_+(\omega,\left|\vec{p}\right|)\Big\{\delta(E-\omega+E_1)
  (1-\hat{\bf q}\cdot\hat{\bf k})f_\mathrm{FD}(-E_1)+\delta(E-\omega-E_1)
  (1+\hat{\bf q}\cdot\hat{\bf k})f_\mathrm{FD}(E_1)\Big\}\nonumber \\ 
  && \quad +\rho_-(\omega,\left|\vec{p}\right|)\Big\{\delta(E-\omega+E_1)
  (1+\hat{\bf q}\cdot\hat{\bf k})f_\mathrm{FD}(-E_1)+\delta(E-\omega-E_1)
  (1-\hat{\bf q}\cdot\hat{\bf k})f_\mathrm{FD}(E_1)\Big\}\Big]
\ee
with $E_1=\left|\vec{q}-\vec{p}\right|$. Since both $E\gg T$ and $E_1\gg T$, 
the terms proportional to $\delta(E-\omega+E_1)$ are exponentially suppressed 
by the factor $f_\mathrm{FD}(\omega)$ for $\omega\gg T$.  
Also, because $E_1$ corresponds to a massless parton, energy and momentum 
conservation does not permit the terms containing $\delta(E+|\omega|-E_1)$ 
for $\omega^2>\left|\vec{q}\right|^2$, i.e.\ there is no phase space available
for the process $1\to 2+3$ if parton $1$ is on-shell. These arguments lead
to further simplifications and we have
\be
  {\rm Im}\, \Pi^{R   \mu}_{ \mu} &=&6e^2 \pi\sum_f\left(\frac{e_f}{e}\right)^2
  \left(1-e^{E/T}\right)\int\frac{d^3q}{(2\pi)^3}\nonumber \\
  && \times \Big[f_\mathrm{FD}(\omega_+)f_\mathrm{FD}(E_1)\frac{\omega_+^2-
  \left|\vec{q} \right|^2}{2m_F^2}\delta(E-\omega_+-E_1)(1+\hat{\bf q}
  \cdot\hat{\bf k}) \nonumber \\
  && \quad +f_\mathrm{FD}(\omega_-)f_\mathrm{FD}(E_1)\frac{\omega_-^2 -
  \left|\vec{q}\right|^2}{2m_F^2}\delta(E-\omega_--E_1)
  (1-\hat{\bf q}\cdot\hat{\bf k})\nonumber \\
  && \quad +\int_{-\infty}^{\infty}d\omega f_\mathrm{FD}(\omega)
  f_\mathrm{FD}(E_1)
  \Big\{\beta_+(\omega,\left|\vec{q}\right|)(1+\hat{\bf q}\cdot\hat{\bf k})
  +\beta_-(\omega,\left|\vec{q}\right|)(1-\hat{\bf q}\cdot\hat{\bf k})\Big\}
  \delta(E-\omega-E_1)  \Big].
\ee

Now we analyze the integral over $d^3q=\left|\vec{q}\right|^2d\left|\vec{q}\right|d\Omega$. When $\left|\vec{q}\right|$ becomes as large as $E_1$, the propagator 
associated with $q$ does not have to be resummed.  If $\left|\vec{q}\right|$ gets much bigger, 
implying $E_1\sim g_sT$, then this is the propagator associated with $E_1$ 
which should be resummed. We have a symmetry relatively to the point $\left|\vec{q}\right|=E_1$, 
such that $\int_0^{\infty}d\left|\vec{q}\right|\rightarrow\int_0^{\infty} d\left|\vec{q}\right| 2\Theta(E_1-\left|\vec{q}\right|)$.  
According to Eq.~(\ref{rate}) the dilepton production rate is thus given by
\be
\label{rate_im}
  \frac{dR^{e^+e^-}}{d^4p}&=&\frac{2\alpha^2}{\pi^2M^2}\sum_f
  \left(\frac{e_f}{e}\right)^2\int_0^{\infty}2\left|\vec{q}\right|^2d\left|\vec{q}\right|\int
  \frac{d\Omega}{(2\pi)^3}\nonumber \\
  && \times \Big[f_\mathrm{FD}(\omega_+)f_\mathrm{FD}(E_1)
  \frac{\omega_+^2-\left|\vec{q}\right|^2}{2m_F^2}
  \delta(E-\omega_+-E_1)(1-\hat{\bf q}\cdot\hat{\bf p}_1)\Theta(E_1-\left|\vec{q}\right|)
  \nonumber \\
  && \quad +f_\mathrm{FD}(\omega_-)f_\mathrm{FD}(E_1)
  \frac{\omega_-^2-\left|\vec{q}\right|^2}{2m_F^2}
  \delta(E-\omega_--E_1)(1+\hat{\bf q}\cdot\hat{\bf p}_1)
  \Theta(E_1-\left|\vec{q}\right|)\nonumber \\
  && \quad +\int_{-\infty}^{\infty}d\omega f_\mathrm{FD}(\omega)
  f_\mathrm{FD}(E_1) \delta(E-\omega-E_1)\Theta(E_1-\left|\vec{q}\right|)\nonumber \\
  && \qquad \times \Big\{\beta_+(\omega,\left|\vec{q}\right|)
  (1-\hat{\bf q}\cdot\hat{\bf p}_1)+\beta_-(\omega,\left|\vec{q}\right|)(1+\hat{\bf q}
  \cdot\hat{\bf p}_1)\Big\}\Big]
\ee
where $\hat{\bf p}_1 = \vec{p}_1/|\vec{p}_1|=-\hat{\bf k}$.

 The dilepton production rate from finite-temperature field theory calculated in this section follows the path presented in Refs.~\cite{KLS91,thoma}.  New to our approach is that we keep the general form of the function $f_{FD}(E_1)$ until the end of the calculation, while this function is integrated out and information about it is finally lost,  in the earlier work. In the next section we will analyze, for the first time, the physical processes behind the production rate in relativistic kinetic theory. This is done so that 
$f_\mathrm{FD}(E_1)$ can be interpreted as the phase-space distribution 
function of an incoming parton of energy $E_1$. Once this is established 
one can apply the usual technique and obtain the production rate involving 
jets by substituting the thermal distribution $f_\mathrm{FD}(E_1)$ by the 
time-dependent jet distribution $f_\mathrm{jet}(E_1,t)$.

\begin{figure}
\centerline{\epsfig{file=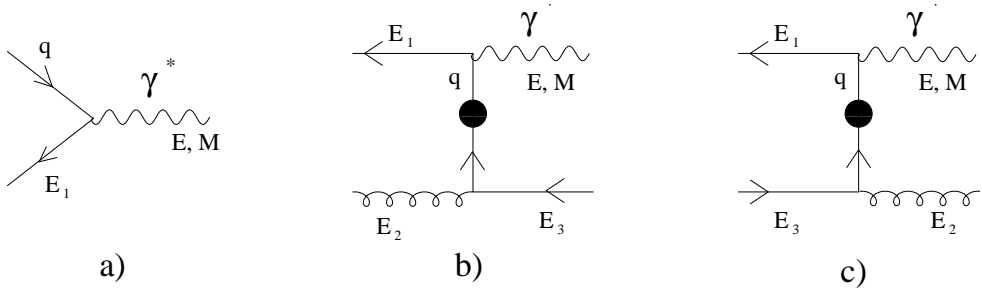,width=13cm}}
\caption{ Processes extracted from photon self-energy in
  Fig.~\ref{pi_gamma_pt}.  In the picture, the particle with energy $E_1$ is 
  an anti-quark, but the quark contribution is also included in the calculation. }
\label{amplitude}
\end{figure}

\section{Dilepton Production Rate from Relativistic Kinetic Theory}
\label{sec:iii}

We will show in this section that the production rate in Eq.~(\ref{rate_im}) 
can also be interpreted in relativistic kinetic theory. Indeed, after cutting 
the diagram in Fig.~\ref{pi_gamma_pt}, we get the Feynman amplitudes shown 
in Fig.~\ref{amplitude}.  The process in Fig.~\ref{amplitude}(a) corresponds 
to the annihilation of the hard antiquark of energy $E_1$ with a soft
quasiparticle $q_+$ or $q_-$. This is the pole-pole contribution, as it involves the propagation of the poles of the propagator in the diagram shown in Fig.~\ref{pi_gamma_pt}.
Fig.~\ref{amplitude}(b) corresponds to
Compton scattering of an antiquark with energy $E_1$ with a hard gluon from the 
medium, with the exchange of a soft quasiparticle and Fig.~\ref{amplitude}(c) 
represents the annihilation of an antiquark $E_1$ with a hard quark from the medium with the exchange of a soft quasiparticle. They correspond to a cut through the self-energy of the dressed propagator in Fig.~\ref{pi_gamma_pt}: this is thus called the cut-pole contribution. There is no
$s$-channel process because hard thermal loops have imaginary parts for 
$q^2<0$ only~\cite{lebellac}. The dots mean that the quasiparticle 
propagator is resummed. That modification of the propagator in the space-like 
region at non-zero temperature is known as Landau damping.  

The production rate, as computed from relativistic kinetic theory (kt), for
reaction $1+2\rightarrow \gamma^*+3+...$ is
\be
  E\frac{dR^{\gamma^*}_{\mbox{kt}}}{d^3p} =
  \int \left( \prod_i \frac{d^3p_i}{2(2\pi)^3E_i} \right)
  (2\pi)^4 \delta^4(p_1+p_2-p-p_3-...) \left|{\cal M}\right|^2
  \frac{f(E_1)f(E_2) (1\pm f(E_3))...}{2(2\pi)^3}
\label{kinetic}
\ee
where ${\cal M}$ is the invariant scattering matrix element.

\subsection{Pole-Pole Contributions ($1+q_{\pm}\rightarrow \gamma^*$)}

The production rate for the process of Fig.~\ref{amplitude}(a) is
\be
  E\frac{dR^{\gamma^*}_{\mbox{kt}a}}{d^3p}&=&\int\frac{d^3p_1}{2(2\pi)^3E_1}
  \frac{d^3q}{2(2\pi)^3q_0} (2\pi)^4\delta^4(p_1+q-p)\left|{\cal M}_{\mbox{kt}a}
  \right|^2\frac{f^{q+\bar{q}}(E_1)f_\mathrm{FD}(q_0)}{2(2\pi)^3}\nonumber \\ 
  &=& \int\frac{d^3q}{8(2\pi)^5q_0E_1} \delta(E_1+q0-E)\left|{\cal M}_{\mbox{kt}a}
  \right|^2 f^{q+\bar{q}}(E_1)f_\mathrm{FD}(q_0)
\ee
where $f^{q+\bar{q}}(E_1)$ is the phase-space distribution of quarks and 
antiquarks with energy $E_1$. The square of the matrix element, after 
summation over spin and color is
\be
  \left|{\cal M}_{\mbox{kt}a}\right|^2 &=&\left|{\cal M}_{1+q_+\rightarrow \gamma^*}
  \right|^2+\left|{\cal M}_{1+q_-\rightarrow \gamma^*}\right|^2\nonumber \\
  &=&6e^2\sum_f\left(\frac{e_f}{e}\right)^2 \mbox{Tr}\left[\not\! p_1
  \left(\sum_s u_s^{q_+}(q)\bar{u}_s^{q_+}(q)+\sum_s u_s^{q_-}(q)
  \bar{u}_s^{q_-}(q)\right)\right]
\label{M_a}
\ee

We need to find a completeness relation for 
$\sum_s u_s^{q_\pm}(q) \bar{u}_s^{q_\pm}(q)$. 
The dressed propagator in Minkowski space is
\be
  -iS_D(q)=\frac{1}{\not\! q -\Sigma}=-\frac{\gamma^0-
  \hat{\gamma}\cdot \hat{\bf q}}{2D_+(q)}-\frac{\gamma^0+
  \hat{\gamma}\cdot \hat{\bf q}}{2D_-(q)}
\label{S_real}
\ee
and by definition, it is~\cite{peskin}
\be
  -iS_D(q)=\frac{\sum_s u_s^{q_+}(q)\bar{u}_s^{q_+}(q)}{q^2-m_{q_+}^2}
  +\frac{\sum_s u_s^{q_-}(q)\bar{u}_s^{q_-}(q)}{q^2-m_{q_-}^2}
\label{S_def}
\ee
where $m_+$ and $m_-$ are the physical mass of the quasiparticles.  
We expand $D_\pm$ around the pole located at $q_0=\omega_\pm$:
\be
  D_\pm(q)&\approx & (q_0-\omega_\pm)\frac{\partial D_\pm}{\partial q_0}
  \Big|_{q_0=\omega_\pm}=(q_0-\omega_\pm)\left(-1+\frac{\partial}{\partial q_0}
  (A\pm B)\right)\Big|_{q_0=\omega_\pm} \nonumber \\ 
  &=& -\frac{m_F^2(q^2-m_\pm^2)}{(\omega_\pm^2-|\vec{q}|^2)\omega_\pm}
\label{residu}
\ee
From Eqs.~(\ref{S_real}), (\ref{S_def}) and (\ref{residu}), we finally obtain
the relation
\be
  \sum_s u_s^{q_\pm}(q)\bar{u}_s^{q_\pm}(q)=\frac{\omega_\pm
  (\omega_\pm^2-|\vec{q}|^2)}{2m_F^2}(\gamma^0\mp\hat{\gamma}\cdot \hat{\bf q})
\ee

Substituting this into Eq.~(\ref{M_a}) gives 
\be
  \left|{\cal M}_{\mbox{kt}a}\right|^2=12e^2\sum_f\left(\frac{e_f}{e}\right)^2
  \frac{E_1}{m_F^2}\left[\omega_+(\omega_+^2-|\vec{q}|^2)(1-\hat{\bf q}
  \cdot\hat{\bf p}_1)+ \omega_-(\omega_-^2-|\vec{q}|^2)(1+\hat{\bf q}
  \cdot\hat{\bf p}_1) \right]
\ee
and the virtual photon production rate becomes
\be
  E\frac{dR^{\gamma^*}_{\mbox{kt}a}}{d^3p}&=& \frac{12}{m_F^2}e^2\sum_f
  \left(\frac{e_f}{e}\right)^2\int\frac{d^3q}{8(2\pi)^5}\nonumber \\ 
  &&\times \Big\{f_\mathrm{FD}(\omega_+)f^{q+\bar{q}}(E_1)(\omega_+^2-\left|\vec{q}\right|^2)
  (1-\hat{\bf q}\cdot\hat{\bf p}_1)\delta(E_1+\omega_+-E)\Theta(E_1-\left|\vec{q}\right|)
  \nonumber \\
  && \quad + f_\mathrm{FD}(\omega_-)f^{q+\bar{q}}(E_1)(\omega_-^2-\left|\vec{q}\right|^2)
  (1+\hat{\bf q}\cdot\hat{\bf p}_1)\delta(E_1+\omega_--E)\Theta(E_1-\left|\vec{q}\right|)\Big\}
\ee
We have introduced here the same cut $\Theta(E_1-\left|\vec{q}\right|)$ as in the previous 
section. Finally the dilepton production rate from the process in 
Fig.~\ref{amplitude}(a) is
\be
  \frac{dR^{e^+e^-}_{\mbox{kt}a}}{d^4p}&=& \frac{2\alpha}{3\pi M^2}E
  \frac{dR^{\gamma^*}_{\mbox{kt}a}}{d^3p}\nonumber \\ 
  &=&\frac{2\alpha^2}{\pi^2M^2}\sum_f\left(\frac{e_f}{e}\right)^2
  \int_0^{\infty}\frac{\left|\vec{q}\right|^2d\left|\vec{q}\right|}{(2\pi)^3}\int d\Omega\nonumber \\ 
  && \times \Big\{f_\mathrm{FD}(\omega_+)f^{q+\bar{q}}(E_1)
  \frac{(\omega_+^2-\left|\vec{q}\right|^2)}{2m_F^2}(1-\hat{\bf q}\cdot\hat{\bf p}_1)
  \delta(E_1+\omega_+-E)\Theta(E_1-\left|\vec{q}\right|)\nonumber \\
  && \quad + f_\mathrm{FD}(\omega_-)f^{q+\bar{q}}(E_1)
  \frac{(\omega_-^2-\left|\vec{q}\right|^2)}{2m_F^2}(1+\hat{\bf q}\cdot\hat{\bf p}_1)
  \delta(E_1+\omega_--E)\Theta(E_1-\left|\vec{q}\right|)\Big\}\, .
\label{rate_a}
\ee

\subsection{Cut-Pole Contributions ($1+2\rightarrow 3+\gamma^*$)}

The expressions for the annihilation and Compton scattering processes from 
Figs.~\ref{amplitude}b and ~\ref{amplitude}c are
\be
  E\frac{dR^{\gamma^*}_{\mbox{kt}bc}}{d^3p}&=&\int\frac{d^3p_1}{2(2\pi)^3E_1}
  \frac{d^3p_2}{2(2\pi)^3E_2}\frac{d^3p_3}{2(2\pi)^3E_3}(2\pi)^4
  \delta^4(p_1+p_2-p-p_3)\left|{\cal M}_{\mbox{kt}b}\right|^2\frac{f^{q+\bar{q}}(E_1)
  f_\mathrm{BE}(E_2)(1- f_\mathrm{FD}(E_3))}{2(2\pi)^3}\nonumber \\ 
  &+&\int\frac{d^3p_1}{2(2\pi)^3E_1}\frac{d^3p_3}{2(2\pi)^3E_3}
  \frac{d^3p_2}{2(2\pi)^3E_2}(2\pi)^4\delta^4(p_1+p_3-p-p_2)
  \left|{\cal M}_{\mbox{kt}c}\right|^2\frac{f^{q+\bar{q}}(E_1)f_\mathrm{FD}(E_3)
  (1+f_\mathrm{BE}(E_2))}{2(2\pi)^3}\nonumber \\
\ee
where $f_\mathrm{BE}$ is the Bose-Einstein distribution function.
The squared matrix elements of the diagrams in Fig.~\ref{amplitude}b and 
Fig.~\ref{amplitude}c are given by
\be
\label{M}
  \left|{\cal M}_{\mbox{kt}b}\right|^2=\left|{\cal M}_{\mbox{kt}c}\right|^2 =
  3 C_F g_s^2 e^2\sum_f\left(\frac{e_f}{e}\right)^2\mbox{Tr}
  \left[4\not\!p_1 S_D^*(p-p_1)\not\!p_3 S_D(p-p_1)   \right]\, ,
\ee
where $C_F=4/3$ is the quark Casimir. It is important to point out that only 
the $t$-channel exchange is considered in the Compton scattering process.  
For the annihilation process shown in Fig.~\ref{amplitude}c, the particle 
with energy $E_1$ is an antiquark and the one with energy $E_3$ is a quark. 
There is also a contribution with $E_1$ and $E_3$ being associated with
a quark and an antiquark respectively. Those two contributions are added 
incoherently, since coherence effect are suppressed by higher powers in
$g_s$. 
After inserting $1=\int_{-\infty}^\infty d\omega\, \delta(\omega-E+E_1)$ 
and using $p_1=p-q$ we obtain
\be
\label{eq_31}
  E\frac{dR^{\gamma^*}_{\mbox{kt}bc}}{d^3p}&=&\int_{-\infty}^\infty d\omega\, 
  \int\frac{d^3q}{16(2\pi)^5E_1}\frac{d^3p_2}{(2\pi)^3E_2E_3}
  \delta(\omega-E+E_1)\nonumber \\ && \times\Big[ \delta(\omega-E_2+E_3)
  \left|{\cal M}_{\mbox{kt}b}\right|^2 f^{q+\bar{q}}(E_1)f_\mathrm{BE}(E_2)
  (1- f_\mathrm{FD}(E_3))\nonumber \\ && \quad +\delta(\omega-E_3+E_2)
  \left|{\cal M}_{\mbox{kt}c}\right|^2 f^{q+\bar{q}}(E_1)f_\mathrm{FD}(E_3)
  (1+f_\mathrm{BE}(E_2))
  \Big.]\nonumber 
\ee

\begin{figure}
\centerline{\epsfig{file=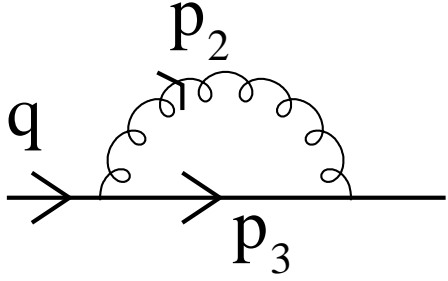,width=6cm}}
\caption{ Quark self-energy with gluon propagator in the loop.}
\label{quark_self}
\end{figure}

To compare this result with the one obtained in the last section
we take advantage of the quark self-energy $\Sigma$, shown in 
Fig.~\ref{quark_self}.  The expression for the discontinuity of $\Sigma$ 
in the space-like region is~\cite{weldon} 
\be
\label{disc_s}
  \mbox{Disc}\, \Sigma(\omega,|\vec{q}|)&=&-i\pi C_Fg_s^2\int
  \frac{d^3p_2}{(2\pi)^3E_2E_3}\not\!p_3\Big\{\delta(\omega-E_3+E_2)
  f_\mathrm{FD}(E_3)(1+f_\mathrm{BE}(E_2))\nonumber \\ 
  && +\delta(\omega+E_3-E_2)f_\mathrm{BE}(E_2)(1-f_\mathrm{FD}(E_3))\Big\}
  f_\mathrm{FD}^{-1}(\omega)
\ee
Eqs.~(\ref{M}), (\ref{eq_31}) and (\ref{disc_s}) lead to
\be
\label{eq_notr}
  E\frac{dR^{\gamma^*}_{\mbox{kt}bc}}{d^3p}&=&3ie^2\sum_f\left(\frac{e_f}{e}
  \right)^2\int_{-\infty}^\infty d\omega\, \int\frac{d^3q}{8(2\pi)^6E_1}
  \delta(\omega-E+E_1)f^{q+\bar{q}}(E_1)f_{FD}(\omega)\nonumber \\ 
  &&\times\, \mbox{Tr}\left[4\not\!p_1 S_D^*(q)\,\mbox{Disc}\,
  \Sigma(\omega,|\vec{q}|)\, S_D(q)\right]
\ee
We can use the relation
\be
  S_D^*(q)\,\mbox{Disc}\,\Sigma(\omega,|\vec{q}|)\, S_D(q)=\mbox{Disc}
  \left(-iS_D(q)\right)
\ee
which holds given that $D_\pm(q)=D_\pm^*(q^*)$.  This is indeed the case as
can be inferred from the definition of $D_\pm$, Eqs.~(\ref{D_def}) and 
(\ref{A_B_def}).
For $\omega^2-|\vec{q}|^2<0$, we use Eqs.~(\ref{rho}), (\ref{rho_2}) and 
(\ref{S_real}) to express the right hand side as
\be
  \mbox{Disc}\,\left(-iS_D(q)\right)
  &=&-\frac{(\gamma^0-\hat{\gamma}\cdot \hat{\bf q})}{2}
  \mbox{Disc}\,\frac{1}{D_+(q)} - \frac{(\gamma^0+
  \hat{\gamma}\cdot \hat{\bf q})}{2}\mbox{Disc}\,\frac{1}{D_-(q)}\nonumber \\
  &=&-i(\gamma^0-\hat{\gamma}\cdot \hat{\bf q})\mbox{Im}\,
  \frac{1}{D_+(q)}-i(\gamma^0+\hat{\gamma}\cdot \hat{\bf q})
  \mbox{Im}\,\frac{1}{D_-(q)}\nonumber \\
  &=&-i\pi(\gamma^0-\hat{\gamma}\cdot \hat{\bf q})\beta_{+}(\omega,
  \left|\vec{q}\right|)-i\pi(\gamma^0+\hat{\gamma}\cdot \hat{\bf q})
  \beta_{-}(\omega,\left|\vec{q}\right|)\, .
\ee

Using the latter result in Eq.~(\ref{eq_notr}) and carrying out the trace, 
we find that
\be
  E\frac{dR^{\gamma^*}_{\mbox{kt}bc}}{d^3p}&=&3e^2\sum_f\left(\frac{e_f}{e}
  \right)^2\int_{-\infty}^\infty d\omega\, 
  \int_0^\infty\frac{\left|\vec{q}\right|^2d\left|\vec{q}\right|}{(2\pi)^5}\int d\Omega\, \delta(\omega-E+E_1)
  f^{q+\bar{q}}(E_1)f_\mathrm{FD}(\omega)\nonumber \\ 
  &&\times\, \left[\beta_{+}(\omega,\left|\vec{q}\right|)(1-\hat{\bf q}
  \cdot\hat{\bf p_1})+\beta_{-}(\omega,\left|\vec{q}\right|)(1+\hat{\bf q}
  \cdot\hat{\bf p_1})\right]\Theta(E_1-\left|\vec{q}\right|)
\ee
As before, we have introduced the term $\Theta(E_1-\left|\vec{q}\right|)$ as we consider only the
region where HTL may be important.  The dilepton pair production rate for the
process shown in Figs.~\ref{amplitude}b and c is
\be
\label{rate_bc}
  \frac{dR^{e^+e^-}_{\mbox{kt}bc}}{d^4p}&=& \frac{2\alpha}{3\pi M^2}E
  \frac{dR^{\gamma^*}_{\mbox{kt}bc}}{d^3p}\nonumber \\ 
  &=&\frac{2\alpha^2}{\pi^2M^2}\sum_f\left(\frac{e_f}{e}\right)^2
  \int_{-\infty}^\infty d\omega\, \int_0^\infty\frac{\left|\vec{q}\right|^2d\left|\vec{q}\right|}{(2\pi)^3}
  \int d\Omega\, \delta(\omega-E+E_1)f^{q+\bar{q}}(E_1)f_\mathrm{FD}(\omega)
  \nonumber \\ 
  &&\times\, \left[\beta_{+}(w,\left|\vec{q}\right|)
  (1-\hat{\bf q}\cdot\hat{\bf p_1})+\beta_{-}(w,\left|\vec{q}\right|)
  (1+\hat{\bf q}\cdot\hat{\bf p_1})\right]\Theta(E_1-\left|\vec{q}\right|)
\ee

Upon adding Eqs.~(\ref{rate_a}) and (\ref{rate_bc}), we reproduce the result 
from Eq.~(\ref{rate_im}), when the particle associated to $E_1$ is thermal, 
i.e $f^{q+\bar{q}}(E_1)\rightarrow 2f_\mathrm{FD}(E_1)$. This proves that 
both methods, finite-temperature field theory and the relativistic kinetic 
formalism, lead to the same result.  

We now briefly compare our approach with the method used by Thoma and Traxler 
in Ref.~\cite{thoma}. They have calculated the photon self-energy shown in 
Fig.~\ref{pi_gamma_pt} with an imposed cutoff $k_s \ll T$ on the momentum
$\left|\vec{q}\right|$ in the loop-integral, such that $0\le \left|\vec{q}\right|\le k_s$. They then added the Compton scattering and annihilation 
processes coming from cutting the two-loop photon self-energy without HTL 
propagators or HTL vertices.  Those two latter process have an infrared 
divergence, which is regulated by imposing a low value cutoff $k_s$ for the 
exchange momentum.  When adding all those processes, the final production rate 
is infrared safe and independent of $k_s$. They have also calculated the 
$\alpha^2\alpha_s$ contribution coming from the pole of the effective quark 
propagator in Fig.~\ref{pi_gamma_pt}. However in their approach, the 
information about the parton phase space distribution is lost, i.e. it is not 
possible at the end to make the substitution $f^{q+\bar{q}}\rightarrow 
f^{q+\bar{q}}_\mathrm{jet}$. Here, we only consider the one-loop 
diagram from Fig.~\ref{pi_gamma_pt}, but we use the dressed propagator 
$S_D(q)$ up to the scale $\left|\vec{q}\right|=k_c$, where $k_c$ corresponds
to $E_1$ due to the $\theta(E_1-\left|\vec{q}\right|)$ function. 
With this method we do not have to specify the shape of $f^{q+\bar{q}}$ until 
the end of the calculation.  We have verified that our numerical result 
depends only weakly on the scale $k_c$. For example, taking $k_c=0.6\times E_1$ reduces the production rate by $\sim$ 20 $\%$.

\begin{figure}
\centerline{\epsfig{file=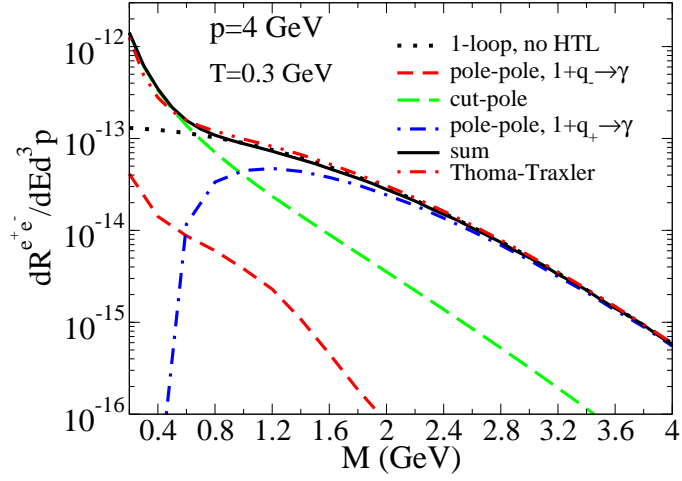,width=7cm,angle=-90}}
\caption{(Color online) Production rate of dileptons with momentum $p=4$ GeV, 
from thermally induced reactions, at a temperature $T=300$ MeV and for $\alpha_s=0.3$. Dotted line: 
Born term; short-dashed line: pole-pole contribution with particles having 
negative identical helicity over chirality ratio $\chi$; dot-dashed line: 
pole-pole contribution with particles having identical helicity over chirality 
ratio $\chi$; long-dashed line: cut-pole contribution; solid line: sum of all 
processes; and double dot-dashed line: Born term plus $\alpha^2\alpha_s$ 
contributions from Ref.~\cite{thoma}. }
\label{htl_vs_thoma}
\end{figure}

Fig~\ref{htl_vs_thoma} shows, for $f^{q+\bar{q}}(E_1)\rightarrow 
2f_\mathrm{FD}(E_1)$, the different sources of dileptons at a temperature 
$T=300$ MeV. In all cases, the particle with energy $E_1$ corresponds to a 
pole with positive $\chi$. The pole-pole contributions are shown by the 
dot-dashed and the short-dashed lines. They correspond to the diagram
in Fig.~\ref{amplitude}a. The annihilation of two partons with positive 
helicity over chirality ratio, $\chi=1$, (dotted-dashed line) dominates at high 
invariant mass. For $M > 1$ GeV it converges toward the Born term 
(dotted line) obtained from a one-loop photon self-energy calculation
with two bare propagators.  
The cut-pole contribution (long-dashed line) is dominant for $M < 1$ GeV.  
The corresponding physical processes are the annihilation and Compton
processes in Fig.~\ref{amplitude}b and c. The sum of our contribution is
shown as the solid line.
It agrees very well (within 30 $\%$) with the sum of the Born term and the 
$\alpha^2\alpha_s$ result from Ref.~\cite{thoma}, given by the double 
dot-dashed line.

\section{Dilepton Yield in Ultra-Relativistic Heavy-Ion Collisions }
\label{sec:iv}

\subsection{Thermal Dileptons}

High-$p_T$ real and virtual photons are preferentially emitted 
early during the QGP phase, when the temperature is largest.
Explicit hydrodynamic calculations show that the
space-time geometry of the fireball smoothly evolves from a 1-D to a 3-D
expansion~\cite{kolb}. 
By the time the system reaches the temperature corresponding
to the mixed phase in a first-order phase transition, the system is still 
dominated by a 1-D expansion~\cite{kolb}. For such a geometry, specific 
calculations~\cite{flow} suggest that the flow effect on 
photons and dileptons from the QGP is not large at RHIC and LHC for
particles with transverse momentum $p_T > 2$ GeV.  

Assuming a 1-D expansion~\cite{Bjorken} that is cut off at
a maximal space-time rapidity $\eta_{\rm max}$, the yield as a function of 
invariant mass $M$ and dilepton rapidity $y_d$
is given by the rate $R^{e^+e^-}=R^{e^+e^-}(\tau,\eta,\mathbf{r}_\perp)$ as
\begin{equation}
\label{dN}
  \frac{dN^{e^+e^-}}{dM^2dy_d} = 
  \int \tau d\tau d^2 r_\perp d\eta \frac{dR^{e^+e^-}}{dM^2 dy_d}\, .
\end{equation}
Here $\tau$, $\eta$ and $\mathbf{r}_\perp$ are the proper time, space-time
rapidity and the transverse coordinate respectively.
In the remainder of this work the dependence of $R$ on $\tau$ and 
$\mathbf{r}_\perp$ will be omitted in the notation for brevity.
As long as $y_d << \eta_{\rm max}$ we can invoke boost invariance for 
$dR^{e^+e^-}/dy_d$ to argue that
\begin{eqnarray}
  \int\limits_{-\eta_{max}}^{\eta_{max}} d\eta \frac{dR^{e^+e^-}(\eta)}{
  dM^2 dy}\Big|_{y=y_d} = 
  \int\limits_{-\eta_{max}}^{\eta_{max}} d\eta \frac{dR^{e^+e^-}(y_d)}{
  dM^2 dy}\Big|_{y=\eta} 
  = \frac{dR^{e^+e^-}(y_d)}{dM^2}.
\end{eqnarray}
Hence the rate of lepton pairs with a fixed rapidity $y_d$ integrated over the
entire longitudinal extent of the fireball is the same as the rate of
all lepton pairs integrated over rapidity from the slice of the fireball at
$\eta=y_d$. In the following we study only dileptons produced at
mid-rapidity $y_d=0$. Their yield can now be written as
\begin{equation}
\label{dn_dm2dyd}
  \frac{dN^{e^+e^-}}{dM^2dy_d}\Big|_{y_d=0}=\int \tau d\tau d^2 r_\perp 
  \frac{dR^{e^+e^-}(\eta=0)}{dM^2}=\int \tau d\tau d^2 r_\perp \int dp_T p_T\int 
  dp_z\frac{2\pi}{E_0}\frac{1}{2}\frac{dR^{e^+e^-}(\eta=0)}{d^4p} \, .
\end{equation}
Here $E_0$ is the energy of the lepton pair in a local frame 
(where the temperature is defined) moving with rapidity $y=\frac{1}{2}\mbox{ln}\frac{E_0+p_0}{E_0-p_0}$ relatively to the fireball.
In this local frame, the transverse and longitudinal momentum of the dilepton are  $p_T$ and $p_z$ respectively. Then we have $p_0=\sqrt{p_T^2+p_z^2}$ and $E_0 = \sqrt{M^2+p_0^2}$. The dilepton
 energy as seen from the fireball is $E=\sqrt{M^2+p_T^2}$.

At this point, we add two constraints in order to facilitate comparison
with experimental data. First, we introduce a lower cutoff 
$p_{T{\rm cut}}$ for the transverse momentum of the lepton pair; 
 second, a cut on the individual lepton rapidities which reflects 
the finite geometrical acceptance of any detector, is included, such that 
$\left|y_{e^{\pm}}\right|\leq y_{\rm cut}$. We introduce a multiplicative 
factor $P_{\rm cut}=P(|y_{e^{\pm}}|\leq y_\mathrm{cut},p_T)$ to include the 
latter condition.
In the center of mass frame of the lepton pair the distribution of positive 
leptons, normalized to unity, is given by
\begin{equation}
  \frac{E^+_{cm}dP_{\gamma^*\to e^+e^-}}{d^3p^+_{cm}}=
  \frac{\delta(E^+_{cm}-\frac{M}{2})}{4\pi E^+_{cm}}.
\end{equation}

Then the probability for a virtual photon with momentum $p_T$ at midrapidity 
to emit two leptons with rapidities $\left|y_{e^{\pm}}\right|\leq y_{\rm cut}$ 
can be obtained by a boost back to the lab frame as
\begin{eqnarray}
\label{prob} 
   P(|y_{e^{\pm}}|\leq y_\mathrm{cut},p_T) 
  &=& \int \frac{d^3p^+_{cm}}{E^+_{cm}} \frac{E^+_{cm}
  dP_{\gamma \to e^+e^-}}{d^3p^+_{cm}}\Theta(|y_{e^\pm}|\le y_{\rm cut})
  \nonumber \\ 
  &=& \int \frac{d\cos\theta d\phi}{\pi} 
   \left(\frac{E^+}{M}\right)^2 \Theta(|y_{e^\pm}|\le y_{\rm cut}).
\end{eqnarray}
Note that in this frame
\begin{eqnarray}
  y_{e^\pm}&=&\frac{1}{2}\ln \frac{E^\pm+p^\pm_z}{E^\pm-p^\pm_z}\, , \\
  E^+ &=& \frac{M^2}{2(\sqrt{M^2+p_T^2}-p_T\cos\theta)}\, , \\
  p_z^\pm &=& \pm E^+\sin\theta\, \sin\phi \,
\end{eqnarray}
and $E^- = E-E^+$. 

The dilepton yield in a longitudinally expanding QGP 
is finally given by
\be
\label{yield_final}
  \frac{dN^{e^+e^-}}{dM^2dy_d}=\int d\tau\, \tau  \int_0^{R_\perp} dr\, r 
  \int_0^{2\pi} d\phi \int_{p_{T_\mathrm{cut}}}^\infty dp_T\, 
  p_T \int_{-\infty}^\infty dp_z\frac{2\pi}{E_0}\frac{1}{2}
  \frac{dR^{e^+e^-}}{d^4p}P(|y_{e^{\pm}}|\leq y_\mathrm{cut},p_T)
\ee 
The thermal-thermal yield at midrapidity is thus obtained by combining
Eq.~(\ref{rate_im}) and Eq.~(\ref{yield_final}).

\subsection{Jet-Thermal Dileptons}

In this subsection, we calculate the emission rate of dileptons from interaction of jets with the medium.
The initial phase space distribution function for partons from jets produced 
in heavy ion collisions is~\cite{simon2,lin}
\be
\label{fjet}
  f_{q\bar{q}}^{\rm jet}(\vec{x},\vec{p},t_0) = \frac{(2\pi)^3
  \mathcal{P}(\mathbf{r}_\perp)}{g_q\tau p_T}
  \frac{dN_{q\bar{q}}^{\rm jet}}{dyd^2p_T}\delta(\eta-y)
\ee
in a boost invariant Bjorken scenario. $\eta$ is the space-time rapidity, 
$t_0$ the formation time of the jets and $\tau^2 = t_0^2-z^2$. 
$dN_{q\bar{q}}^{\rm jet}/(dyd^2p_T)$ is the spectrum of jets and $g_q$ is 
the spin-color degeneracy.  $\mathcal{P}(\mathbf{r}_\perp)$ represents the 
probability to create a jet at position $\mathbf{r}_\perp$ in the transverse 
plane. It is given by the product of the thickness functions of the colliding 
nuclei at $\mathbf{r}_\perp$ normalized to 1. For central collisions and 
assuming hard sphere nuclei, we have
\begin{equation}
  \mathcal{P}(\mathbf{r}_\perp)= {2\over \pi R_\perp^2}
  \left(1 - {r_\perp^2\over R_\perp^2}\right)
  \, \theta(R_\perp - r_\perp) \, ,
\label{Pr_dep}
\end{equation}
where $R_\perp=A^{1/3} 1.2$ fm is the radius of the nucleus in the transverse
plane. Since energy loss by bremsstrahlung involves only small $O(\gs T/p)$ 
changes to the directions of particles, we suppose that jets keep propagating 
in straight lines after they are created.  Then at any later time $t$
\begin{equation}
\label{fjet2}
  f_{q\bar{q}}^{\rm jet}(\vec{x},\vec{p},t) = f_{q\bar{q}}^{\rm jet}
  \left( \vec{x}-\hat{t} \frac{\vec{p}}{\left|\vec{p}\right|},
  \vec{p},t_0 \right)
\end{equation}
where $\hat{t}=t-t_0$ is the propagation time of the jet.  

Energy loss can be described as a dependence of the parton spectrum 
$dN_{q\bar{q}}^{\rm jet}/dE$ on time. We will only be interested in the 
region around midrapidity later, so we can write
\begin{equation}
  \frac{dN_{q\bar{q}}^{\rm jet}}{dy d^2p_T}\Big|_{y=0} = 
  \Omega^{-1} \frac{dN_{q\bar{q}}^{\rm jet}}{E d E}
\end{equation}
with a phase space factor $\Omega$. The evolution of the energy spectrum
$dN_{q\bar{q}}^{\rm jet}(t)/dE$ is determined by the AMY evolution equations.
Note that we do not need to specify the constant $\Omega$ to obtain the 
evolution of the spectrum. Here we quote only the basic formulas of the AMY
formalism, which assumes that jets lose energy by inelastic processes only.
More details can be found in \cite{AMY,Jeon-Moore}.
This formalism has been applied in Ref.~\cite{simon2} to successfully reproduce the $\pi^0$ spectra at RHIC.  The parton spectrum $dN_{q\bar{q}}^{\rm jet}/dE$ depends on the energy lost, but not really on how this
energy has been lost, such that using an another model for jet-quenching, 
like for example a model including elastic energy loss~\cite{whdg05},  should
not affect our following results for the dilepton production from jet-plasma interactions.  
 
The evolution equations for the spectra are
\begin{widetext}
\begin{eqnarray}
  \frac{d}{dt}\left(\frac{dN^{\rm jet}_{q\bar{q}}(p)}{dE}\right) 
  & = & \int dk \! \left[ \frac{dN^{\rm jet}_{q\bar{q}}(p{+}k)}{dE}  
  \frac{d\Gamma^q_{\!qg}(p{+}k,k)}{dkdt} -\frac{dN^{\rm
  jet}_{q\bar{q}}(p)}{dE} \frac{d\Gamma^q_{\!qg}(p,k)}{dkdt} 
  +2 \frac{dN^{jet}_g(p{+}k)}{dE} \frac{d\Gamma^g_{\!q \bar q}(p{+}k,k)}{dkdt}
  \right] \,  , \nonumber \\
  \frac{d}{dt}\left(\frac{dN^{\rm jet}_g(p)}{dE}\right) 
  & = & \int dk \! \left[ \frac{dN^{\rm jet}_{q\bar{q}}(p{+}k)}{dE}  
  \frac{d\Gamma^q_{\!qg}(p{+}k,p)}{dkdt} {+}\frac{dN^{\rm jet}_g(p{+}k)}{dE} 
  \frac{d\Gamma^g_{\!\!gg}(p{+}k,k)}{dkdt} \right. \nonumber \\ 
  && \; \left. -\frac{dN^{\rm jet}_g(p)}{dE} 
  \left(\frac{d\Gamma^g_{\!q \bar q}(p,k)}{dkdt}
  + \frac{d\Gamma^g_{\!\!gg}(p,k)}{dkdt} \Theta(2k{-}p) \!\!\right) \right] ,
\label{eq:Fokker}
\end{eqnarray}
with the transition rates
\begin{equation}
\label{eq:dGamma}
\frac{d\Gamma(p,k)}{dk dt}  =  \frac{C_s \gs^2}{16\pi p^7}
        \frac{1}{1 \pm e^{-k/T}} \frac{1}{1 \pm e^{-(p-k)/T}}
\times
\left\{ \begin{array}{cc}
        \frac{1+(1{-}x)^2}{x^3(1{-}x)^2} & (q \rightarrow qg) \\
        N_{\rm f} \frac{x^2+(1{-}x)^2}{x^2(1{-}x)^2} & (g \rightarrow qq) \\
        \frac{1+x^4+(1{-}x)^4}{x^3(1{-}x)^3} & (g \rightarrow gg) \\
        \end{array} \right\} \times   \int \frac{d^2 \h}{(2\pi)^2} 2 \h \cdot {\rm Re}\: \F(\h,p,k) \, .
\end{equation}
\end{widetext}
The integration over the range $k<0$ represents absorption of
thermal gluons from the QGP; the range with $k>p$ represents
annihilation with a parton from the QGP of energy $k{-}p$, while $0<k<p$
is the range of bremsstrahlung. In writing (\ref{eq:Fokker}), we used
$d\Gamma^g_{\!gg}(p,k)=d\Gamma^g_{\!gg}(p,p{-}k)$ and similarly for
$g\rightarrow qq$. The $\Theta$-function in the loss term for $g
\rightarrow gg$ prevents the double counting of final states.  
In Eq.\ (\ref{eq:dGamma}) $C_s$ is the quadratic Casimir relevant for each 
process considered, and $x\equiv k/p$ is the momentum fraction of the gluon 
(or the quark, for the case $g \rightarrow q\bar{q}$).
The factors $1/(1\pm e^{-k/T})$ are Bose enhancement or Pauli blocking
factors for the final states. The vector $\h \equiv \p \times \mathbf{k}$ determines 
how non-collinear the final state is, and $\F(\h,p,k)$ is the solution of 
an integral equation describing how $\left|p-k;k\right>\left<p\right|$ 
evolves with time~\cite{AMY}.

The dileptons produced from the passage of jets through the QGP are finally 
obtained from Eq.~(\ref{yield_final}) and from Eqs.~(\ref{rate_a}) and 
(\ref{rate_bc}) with the substitution $f^{q+\bar{q}}(E_1)\rightarrow 
f_{q\bar{q}}^{\rm jet}(E_1)$.
Note that $\eta=0$ together with the boost invariance of the jet distribution 
imply that the longitudinal momentum of the jet parton vanishes. After some 
algebra, we get
\be
\label{full_eq}
  \frac{dN_\mathrm{jet-th}}{dM^2dy_d} &=&\frac{4\alpha^2\sum_f
  \left(\frac{e_f}{e} \right)^2}{\pi M^2 g_q}\int dt \int_0^{R_\perp} r dr\, 
  \xi\int_{p_{T_\mathrm{cut}}}^\infty dp_{T} 
  P(|y_{e^{\pm}}|\leq y_\mathrm{cut},
  p_T) \int_0^\infty dq_{T}\int_{-\infty}^\infty dq_{z}\frac{1}{E_0}\nonumber\\
  && \times \sum_{j=\pm}\Big[\frac{f_\mathrm{FD}(\omega_j(\left|\vec{q}\right|))}{2m_F^2}
  \frac{(\omega_j(\left|\vec{q}\right|)^2-\left|\vec{q}\right|^2)}{\sqrt{1-\mbox{cos}^2\theta_j}}(1+j\frac{\left|\vec{q}\right|}{E_1^j}-j
  \frac{{\bf q}\cdot{\bf p}}{\left|\vec{q}\right|E^j_1})E^j_1
  \frac{dN_{q\bar{q}}^\mathrm{jet}(t)}{dy_1d^2p_{T_1}}\Big|_{y_1=0}\nonumber \\
  && \quad +\int_{-\infty}^{\infty}d\omega f_\mathrm{FD}(\omega)
  \beta_j(\omega,\left|\vec{q}\right|) \left(1+j\frac{\left|\vec{q}\right|}{E_1}-j\frac{{\bf q}\cdot{\bf p}}{\left|\vec{q}\right|E_1}
  \right) \frac{E_1}{\sqrt{1-\mbox{cos}^2\theta}}
  \frac{dN_{q\bar{q}}^\mathrm{jet}(t)}{dy_1d^2p_{T_1}}\Big|_{y_1=0}\Big]
\ee
with $\left|\vec{q}\right|=\sqrt{q_T^2+q_z^2}$ and
\be
\xi=\left\{
\begin{array}{l}
  0  \\
  \displaystyle
  \frac{4}{R_\perp^2}
  \left(1-\frac{r^2+ t^2}{R_{\perp}^2}\right)
  \\
  \displaystyle
  \frac{4 u_0}{\pi R_{\perp}^2}\left(1-\frac{r^2+ t^2}{R_{\perp}^2}\right)
  +\frac{8 tr}{\pi R_{\perp}^4}\sin u_0 \,,
\end{array}
\right.
\ee
for the cases that $r^2+ t^2-2  t r > R_{\perp}^2$, 
$r^2+ t^2+2 tr \le R_{\perp}^2$ and all other cases, respectively.
Here we have defined
\begin{equation}
  u_0= \arccos \frac{r^2+ t^2-R_\perp^2}{2 t\,r}\,.
\end{equation}
The other quantities to be specified are
\be
  {\bf q}\cdot{\bf p}&=&p_T\, q_{T}\, \mbox{cos}\,\theta_\pm +q_{z}^2\, ,
  \nonumber \\
  \theta_\pm&=& \mbox{cos}^{-1}\left(\frac{p_T^2+q^2_{T}-
  (E_0-\omega_\pm)^2}{2p_Tq_{T}}  \right)\, , \\
  E_1^\pm&=&E_0-\omega_\pm\, . \nonumber
\ee

We obtain the jet-thermal dilepton yield without HTL effects, i.e the Born 
term, by keeping only the pole-pole($q_+$) contribution in Eq.~(\ref{full_eq}).
In this case we have to substitute:
\begin{align}
  \beta_\pm &\rightarrow  0\, , \qquad &
  \frac{(\omega_-^2-\left|\vec{q}\right|^2)}{2m_F^2} &\rightarrow  0 \, , \\  
  \frac{(\omega_+^2-\left|\vec{q}\right|^2)}{2m_F^2} &\rightarrow  1 \, , \qquad  &  
  (1+\frac{|\vec{q}|}{E_1^+}-\frac{{\bf q}\cdot{\bf p}}{|\vec{q}|E^+_1})
  &\rightarrow  \frac{M^2}{2E_1E_q}\, , \\
  \omega_+ &\rightarrow  E_q=|\vec{q}|\,.  &
\end{align}
This leads to the final expression for the jet-thermal dilepton yield
without HTL effects: 
\be
\label{dn_jetth}
  \frac{dN_\mathrm{jet-th}^\mathrm{no-HTL}}{dM^2dy_d} 
  &=&\frac{2\alpha^2\sum_f\left(\frac{e_f}{e}\right)^2}{\pi g_q}\int dt 
  \int_0^{R_\perp} r dr\, \xi \int_{p_{T_\mathrm{cut}}}^\infty 
  dp_{T}\int_0^\infty dq_{T}\int_{-\infty}^\infty dq_{z}
  \frac{dN^{q\bar{q}}(t)}{dy_1d^2p_{T_1}}\Big|_{y_1=0}\nonumber \\ 
  && \times \frac{1}{E_0E_q\sqrt{1-\mbox{cos}^2\theta_+}}\, 
  f_\mathrm{FD}(E_q)P(|y_{e^{\pm}}|\leq y_\mathrm{cut},p_T)\, .
\ee

For a purely longitudinal
expansion of the fireball~\cite{Bjorken}, at each point
the temperature is evolving from some initial time $\tau_i$ as
\begin{equation}
\label{T1}
  T(\mathbf{r}_\perp,\tau) =T(\mathbf{r}_\perp,\tau_i) 
  \left(\frac{\tau_i}{\tau}\right)^{1/3} \, .
\end{equation}
We assign the initial temperatures in the transverse direction according
to the local density so that~\cite{prlphoton,prcdilep,simon2}
\begin{equation}
\label{T2}
  T(r_\perp,\tau_i)=T_i\left[2\left(1-
  \frac{r_\perp^2}{R_{\perp}^2}\right)\right]^{1/4} \, .
\end{equation}
We assume a first-order phase transition and use
\begin{equation}
f_{\rm QGP}(\tau)=\frac{1}{r_d-1}\left(\frac{r_d\tau_f}{\tau}-1 \right)
\label{fmix}
\end{equation}
as the fraction of the QGP present during the mixed phase~\cite{Bjorken}.
Here $r_d=g_Q/g_H$ is the ratio of the degrees of freedom in the
two phases: $g_Q=42.25$ for a QGP with three flavors of quarks and $g_H=3$ for a simple gas of pions. $\tau_f$ is the time when the temperature reaches the critical temperature of 160 MeV (see Eqs.~\ref{T1} and ~\ref{T2}), marking the beginning of the mixed phase, and 
$\tau_H= r_d \tau_f$ is the time it ends, determined by the condition 
$f_{\rm QGP} = 0$. However, since signals associated with jets are sensitive to early times, the order of the phase transition is not crucial.
The $\tau$-integration 
in (\ref{yield_final}) is carried out from $
\tau_i$ to $\tau_H$ and in addition it is scaled between $\tau_f$ and 
$\tau_H$ to account for the fact that only a fraction of the system is 
still a QGP, such that~\cite{simon2}
\begin{equation}
\label{int_tau}
  \int d\tau = \int\limits_{\tau_i}^{\tau_f}d\tau 
  +\int\limits_{\tau_f}^{\tau_H} d\tau f_{\rm QGP}(\tau)\, .
\end{equation}

\section{Drell-Yan and Heavy Flavour Decay}
\label{sec:v}

We calculate the Drell-Yan process to order $\mathcal{O}(\alpha_s)$ 
in the strong coupling which is the leading order result with 
non-vanishing transverse momentum $p_T$ of the lepton
pair. We also have to take into account virtual photon bremsstrahlung from
jets. The total Drell-Yan yield is the sum of the direct and
the Bremsstrahlung contributions, $\sigma_{\rm DY} = \sigma_{\rm direct} 
+ \sigma_{\rm frag}$~\cite{qiu,berger}.

The direct contribution in collisions of two nuclei $A$ and $B$ is given 
by
\begin{eqnarray}
  \frac{d\sigma_{\rm direct}}{dM^2dy_d\, dp_T^2}=\frac{\alpha^2\alpha_s}{3M^2
    s_{NN}}\sum_{a,b}  \int\frac{dx_a}{x_ax_b}f_{a/A}(x_a,Q) 
   f_{b/B}(x_b,Q) 
  \frac{\left|\bar{M}_{a+b\rightarrow c+\gamma^*}  
  \right|^2}{s_{NN}x_a-\sqrt{s_{NN}}\sqrt{ M^2+p_T^2}e^{y_d}}\, .
\end{eqnarray}
The squared scattering amplitudes $\left|\bar{M}_{a+b\rightarrow c+\gamma^*}  
\right|^2$ for the Compton and annihilation processes of two incoming 
partons can be found in~\cite{guo}. When $p_T$ and $M$ are both large and of the same order, the direct 
contribution is the dominant mechanism.  However, when $\Lambda_{\rm QCD} \ll M
\ll p_T$,
logarithmic corrections with powers of $\ln(p_T^2/M^2)$ are large. They can
be effectively resummed 
into a virtual-photon fragmentation function $D_{\gamma^*/c}(z,Q_F)$, giving
rise to the fragmentation contribution
\begin{eqnarray}
  \frac{d\sigma_{\rm frag}}{dM^2dy_d\, dp_T^2} &=& \frac{\alpha}{3\pi M^2}\int
  \frac{dz}{z^2} \frac{d\sigma^{A+B\rightarrow c+d}}{dy_d\, dp_{cT}^2}
  \Big|_{p_{cT}=p_T/z} \nonumber \\ 
  && \times D_{\gamma^*/c}(z,Q_F)\, .
\end{eqnarray}
The cross sections for the production of a massless partons $c$ in
$A+B$ collisions, $d\sigma^{A+B\rightarrow c+d}/dy_d\, dp_{cT}^2$,
can be found in \cite{owens}.   The factorization scale $Q$ and the 
fragmentation scale $Q_F$ are both set to the order of the 
energy exchanged in the reaction $\sqrt{M^2+p_T^2}$~\cite{yellow_fai}.  
The effect of varying the 
scale $Q= k\sqrt{p_T^2+M^2}$, from $k=1/2$ to $k=2$, introduces a variation
 of the dilepton yield by $\sim\pm$ 35 \% relatively to $k=1$.

The typical fragmentation time of a jet into a virtual photon with large invariant mass $M$ scales like  $1/M$.  Thus dilepton with large mass $M$ can be created in the medium with rather small corrections due to 
energy loss of their parent jet, while low-$M$
dileptons should be formed outside the medium with their yield affected by 
the full energy loss suffered by the jet.
Since the interesting region for the fragmentation process is for low masses
 $M$, where it is expected to be as important as the direct 
process, we assume that virtual photons fragment outside the medium after  
the parent jet has obtained its final energy.
We define a medium-modified effective fragmentation function~\cite{simon2}
\begin{equation}
  D_{\gamma^*/c}(z,Q_F) =\int d^2r_\perp \mathcal{P}(\mathbf{r}_\perp) 
  \int dE_f\frac{z'}{z}\frac{dN^{\rm jet}_{q\bar{q}}}{dE}(E_f;E_i) 
  D^0_{\gamma^*/q}(z',Q_F)
\label{eq:Dtilde}
\end{equation}
where $z=p_T/E_i$ and $z'=p_T/E_f$.  $dN^{\rm jet}_{q\bar{q}}/dE$ is the 
probability to for a given quark with final energy $E_f$ when the initial 
jet is a particle of type $c$ and energy $E_i$, given by the solutions 
to Eq.~(\ref{eq:Fokker}),
$D^0_{\gamma^*/q}$ is the leading order vacuum fragmentation function 
from~\cite{berger,qiu}. The factor $\mathcal{P}(\mathbf{r}_\perp)$ is 
introduced to take care of the spatial distribution of jets and their 
propagation length in the medium.

We implement our cuts for the Drell-Yan process as well, so that the final
yield is given by
\begin{eqnarray}
\label{D-Y}
  \frac{dN_{\rm DY}}{dM^2dy_d} &=& 
  \frac{2\langle N_{\rm coll}\rangle}{\sigma_{in}}\int\limits_{p_{T{\rm
  cut}}}^{\infty}dp_T\, p_T\Big[K_{dir}\frac{d\sigma_{\rm direct}}{dM^2dy_d\, 
  dp_T^2}+K_{frag}\frac{d\sigma_{\rm frag}}{dM^2dy_d\, dp_T^2}\Big]
  P(|y_{e^{\pm}}|\leq y_{cut},p_T)\, .
\end{eqnarray}
where $K_{dir}$ and $K_{frag}$ are introduced to take care of higher order effects. We reproduce the Drell-Yan isolated muons from $\sqrt{s_{NN}}$=630 GeV $p+\bar{p}$ data at CERN~\cite{UA1}, for low invariant masses ($M< 2.5$ GeV) and high transverse momentum ($p_T > 6$ GeV), with a constant factor $K_{dir}=1.5$.  We also take $K_{frag}=1.5$, quite in line with the $K$-factor used for real photons from jet-fragmentation in Ref.~\cite{simon2}.  While we assume the $K$-factors to be constant, they should be in principle $\sqrt{s_{NN}}$, $p_T$ and $M$ dependent.  However, an estimate of the variation of $K_{dir}$ with those parameters can be found in Ref.~\cite{berger}.  For example, next-leading order effects for the direct Drell-Yan contribution at $\sqrt{s_{NN}}=2$ TeV in the range $10< p_T< 20$ GeV would vary within 10$\%$ from $M\sim 2.5$ GeV to $M\sim 4.5$ GeV.

Here we assume $\langle N_{\rm coll}\rangle=975$, $\sigma_{in}=40$ mb for
RHIC~\cite{jeon} and $\langle N_{\rm coll}\rangle=1670$, $\sigma_{in}=72$ 
mb for the LHC~\cite{yellow}.  
In our calculations we use CTEQ5 parton distribution functions \cite{cteq5} 
with EKS98 shadowing corrections~\cite{eskola}.

The main background at RHIC energies for the dilepton production processes 
considered so far is expected to be decay of open charm and bottom mesons
\cite{ralf}.  During the initial hard scattering, $c\, \bar{c}$ ($b\bar b$) 
pairs are produced and can thereafter fragment into $D(B)$ and 
$\bar{D}(\bar{B})$ mesons. We consider here only
correlated decay, which happens when a positron coming from the semileptonic
decay of a $D(B)$ is measured together with the electron from the  
semileptonic decay of a $\bar{D}(\bar{B})$.  The results for heavy
quark decay have been obtained with the techniques of Ref.~\cite{mangano}. 
Collisional energy loss of the heavy quarks propagating in the 
QGP~\cite{heavy_q} has not been included. Since this constitutes a background to our process, we adopt this conservative point of view.

\section{Results}
\label{sec:vi}

We choose the same parameterization of the plasma phase that was previously 
used successfully in the studies of high~\cite{simon2} and low to 
intermediate $p_T$ photons~\cite{simon}. For central Au+Au collisions at 
RHIC we have an initial temperature $T_i$=370 MeV and initial time
$\tau_i$=0.26 fm/c for the plasma phase, corresponding to the particle
rapidity density $dN/dy$=1260~\cite{prcdilep}.  This latest value is obtained
from the measured pseudo-rapidity distribution of charged particles $dN_{ch}/d\eta$~\cite{phobos}: $dN/dy\sim \frac{3}{2}|d\eta/dy|dN_{ch}/d\eta$, where $|d\eta/dy|\sim$ 1.2 around $y=0$.
For LHC, our initial conditions are $T_i$=845 MeV and $\tau_i$=0.088 fm/c corresponding to 
$dN/dy$=5625~\cite{prcdilep,kms}. For the processes involving the QGP, 
we assume three light flavors and we fix $\alpha_s=0.3$.  As the jets are defined to be particles having a transverse momentum 
greater than a scale $p_Q$, with $p_Q\gg 1$GeV, we have set the dilepton 
momentum cutoff $p_{T_\mathrm{cut}}$ high enough to avoid any sensitivity to 
the choice of $p_Q$.  We take $p_{T_\mathrm{cut}}=$ 4(8) GeV for RHIC (LHC). 
The cuts on leptons rapidities emulate the PHENIX experiment at RHIC, 
$y_\mathrm{cut}$=0.35, while we use $y_{cut}=$0.5 at LHC. 

\begin{figure}
\centerline{\epsfig{file=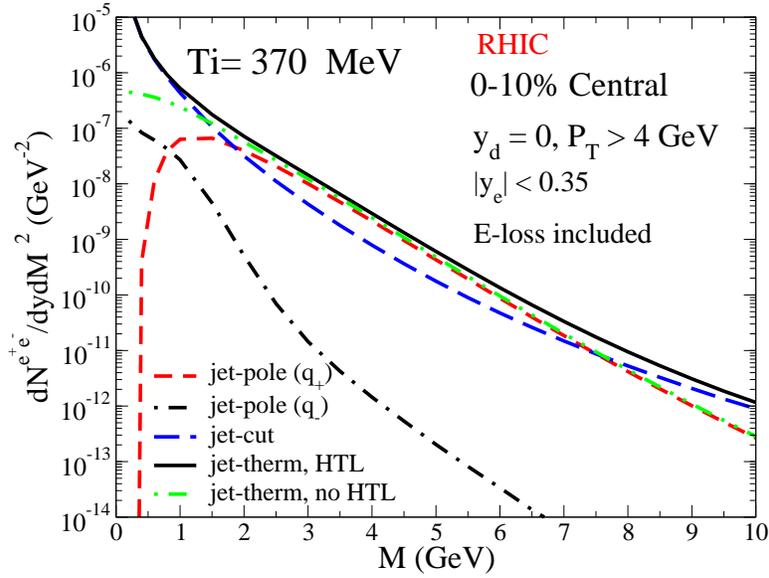,width=8cm,angle=-90}}
\caption{(Color online) High-momentum dileptons produced from the interaction of jets with QGP in Au+Au collisions at 200 GeV (RHIC).  The initial temperature is $T_i$=370 MeV.  Short-dashed line: interaction of jets with poles with positive $\chi$; dot-dashed line: interaction of jets with poles with negative $\chi$; long-dashed line: interaction of jets with cuts; solid line: sum of the latter processes; and double dot-dashed line: interaction of jet with QGP without HTL effects (Born term). See text for details.}
\label{htl_jet}
\end{figure}

The dileptons produced at RHIC by the interaction of jets with the medium are 
shown in Fig.~\ref{htl_jet}. Dileptons from jet-pole interactions, i.e.\
from annihilation of a jet parton with a $(q_-)$-mode, are negligible, 
while the jet-pole interactions involving $(q_+)$-modes tends toward the 
Born term at high invariant mass, as it was the case for thermal-thermal 
reactions in Fig.~\ref{htl_vs_thoma}. 
On the other hand dileptons from jet-cut interactions do not behave like the 
cut-pole process in Fig.~\ref{htl_vs_thoma}. They become 
the dominant contribution at high-$M$. The expressions for the cut-pole 
process, shown in Fig.~\ref{amplitude}b+c, involve the functions 
$\beta_\pm(\left|\vec{q}\right|)$ with $\beta_\pm(\left|\vec{q}\right|)\rightarrow 0$ for $\left|\vec{q}\right|\gg T$.  
When $M$ is large, in order to keep the value of $\left|\vec{q}\right|$ modest, 
the energy $E_1$ of the incoming parton has to be large: $E_1 \ge M^2/4\left|\vec{q}\right|$.  
As the thermal phase space distribution function $f_{FD}$ decreases 
exponentially for large $E_1$, the cut-pole contribution turns out to be 
negligible for large $M$. However, when the incoming particle is a jet with a 
power-law distribution, high values of $E_1$ are not suppressed and the 
cut-pole contribution is important.
Therefore the sum of all the jet-thermal processes with HTL effects included
(solid line), is more important than the jet-thermal contribution without HTL 
effects (double dot-dashed line) by more than a factor 2 for $M$ above 8 GeV.  
For $M$ below 1 GeV HTL corrections increase the Born term by one order of 
magnitude, because of the $1/M^2$ behavior in Eq.~(\ref{full_eq}).  We have 
verified that when extrapolated to $M=0$, the cut-pole contribution 
reproduces the result for the jet-photon conversion in the QGP~\cite{simon2}, 
such that $E^{\gamma^*}dR_{\mbox{cut-pole}}(p,M\to 0)/d^3p=E^{\gamma}dR_{\mbox{jet-th}}(p)/d^3p$.

\begin{figure}[ht!]
\epsfig{file=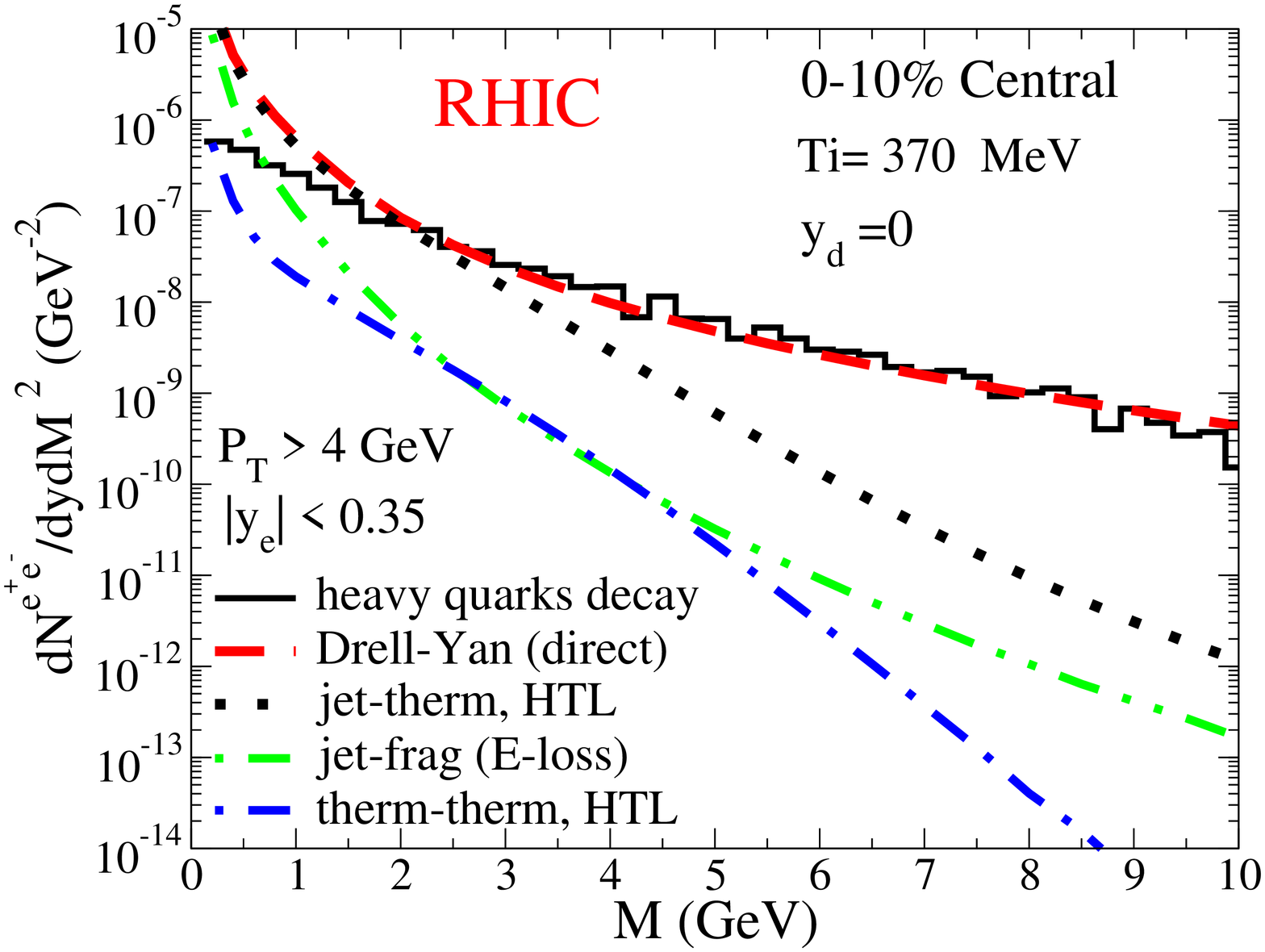,width=6.2cm,angle=-90}
\epsfig{file=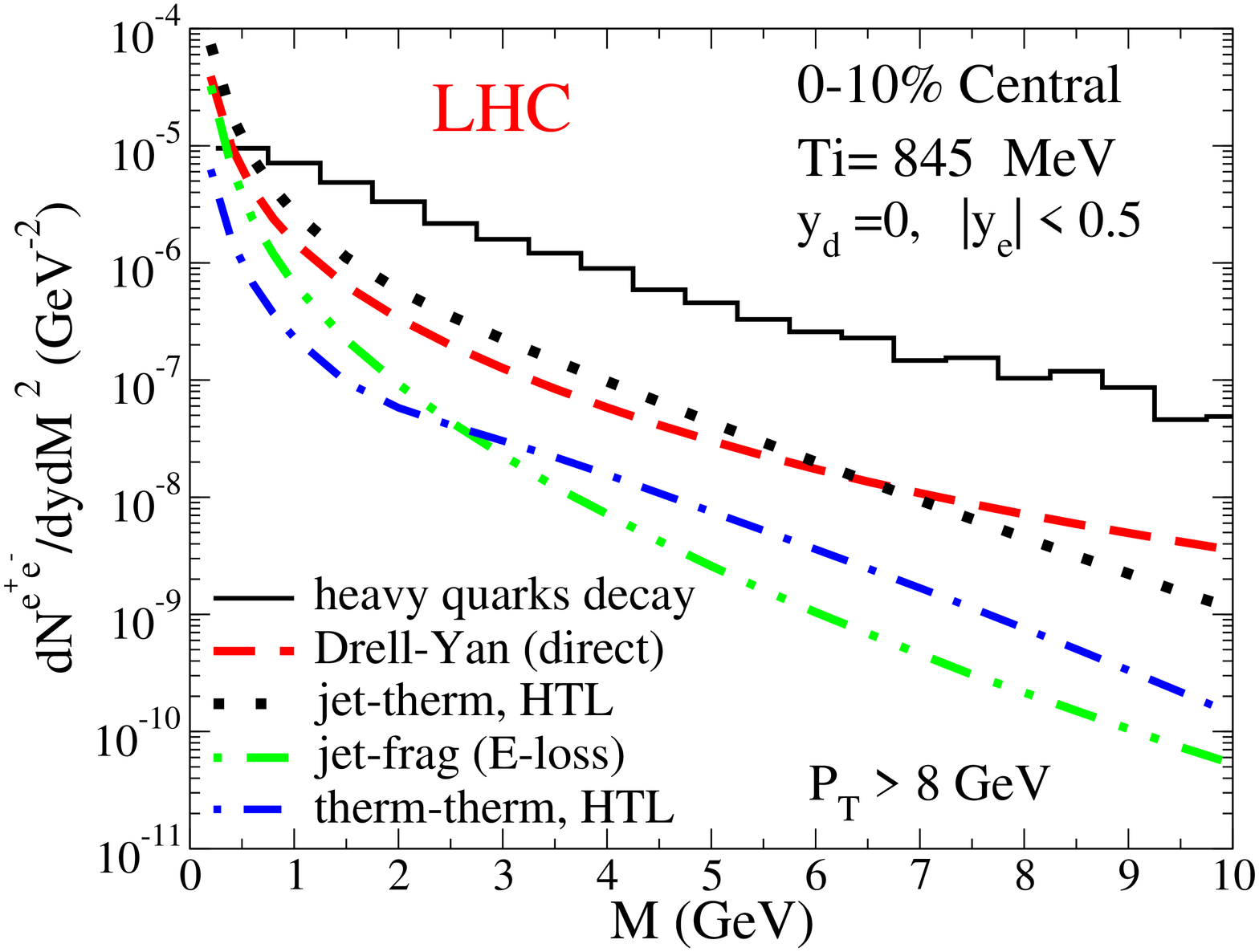,width=6.2cm,angle=-90}
\caption{(Color online)  Sources of high-$p_T$ dileptons in central Au+Au
  collisions at RHIC (left) and Pb+Pb at the LHC (right).  Solid lines:
  semileptonic decay of heavy quarks; dashed lines: direct Drell-Yan 
  contribution; dotted lines: jet-thermal interaction with HTL effects; 
  double dot-dashed lines: jet-fragmentation process; and dot-dashed lines:
  thermal dilepton production with HTL effects.}
\label{compare_dm}
\end{figure}

\begin{figure}
\centerline{\epsfig{file=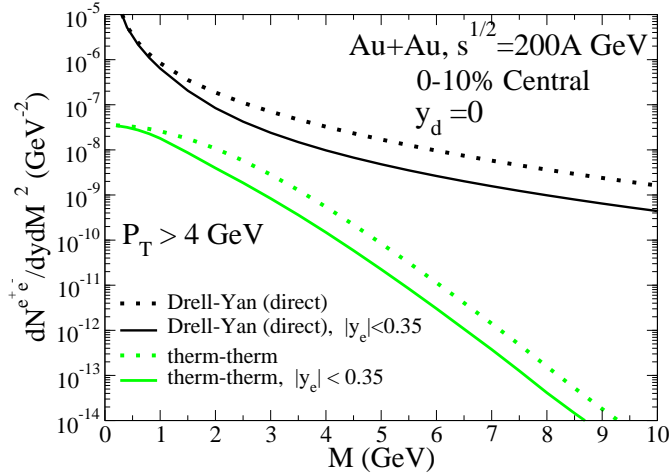,width=7cm,angle=-90}}
\caption{(Color online) Effect of the cut on leptons rapidities for direct 
Drell-Yan (top) and thermal dileptons (bottom) without HTL effects at RHIC.  
The dotted lines include all leptons from the respective process while the 
solid lines include only leptons having absolute value of rapidity smaller 
than 0.35.}
\label{y_cut}
\end{figure}

\begin{figure}
\centerline{\epsfig{file=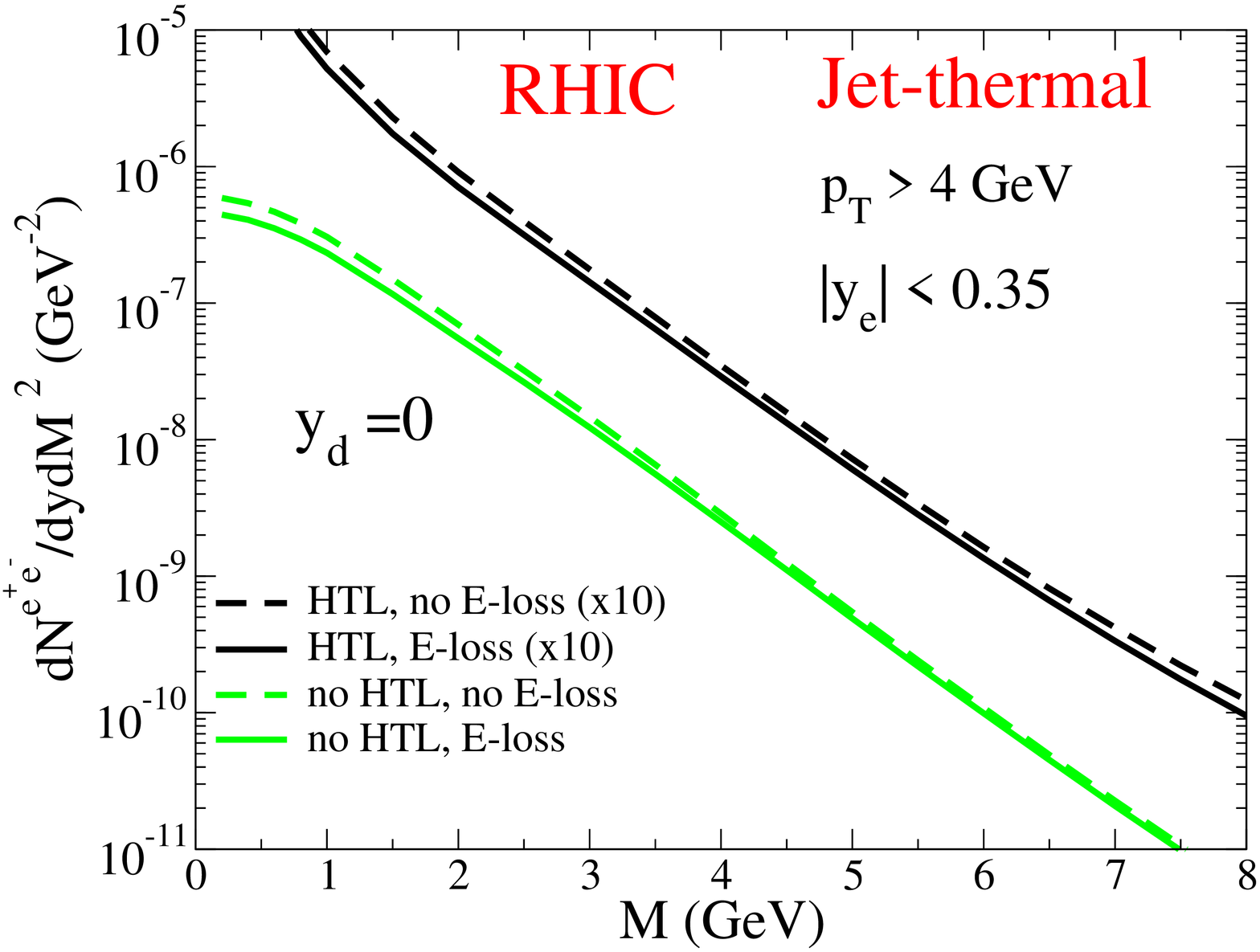,width=7cm,angle=-90}}
\caption{(Color online) Effect of jet energy loss in jet-thermal production of dileptons at RHIC, with HTL effects (top) and without (bottom). The solid lines include the effect of energy loss, while the dashed lines don't.  The results for jet-plasma interactions with HTL have been rescaled by a factor 10 for clarity. }
\label{htl_jet_eloss}
\end{figure}

In Fig.~\ref{compare_dm} we show the mass spectrum of dileptons in
central Au+Au collisions at RHIC ($\sqrt{s_{NN}} = 200$ GeV) and central
Pb+Pb collisions at LHC ($\sqrt{s_{NN}} = 5.5$ TeV).  
For both collider energies, the jet-thermal contribution exceeds the thermal 
dilepton production by an order of magnitude, which was also  
the case for high-$p_T$ photon production~\cite{simon2,prlphoton}.  
However, at RHIC the dominant sources of dileptons for $M>3$ GeV are heavy
quark decay and the direct Drell-Yan process.  At intermediate mass, between 1 GeV and 3 GeV, the jet-thermal contribution 
is as important as these two processes.  Below 1 GeV, the 
fragmentation of jets 
into virtual photons turns out to be 
comparable to the direct Drell-Yan production and the jet-thermal contribution.
At LHC, the whole range of invariant mass is dominated by charm decay, but
on the other hand the jet-thermal lepton pairs exceed the direct Drell-Yan
yield below 7 GeV. We want to stress that energy loss
of heavy partons has not been included here, so that the heavy quark 
contribution is likely an upper limit of what is to be expected.
Lepton pair production from jet-plasma interactions being as important as 
 this upper limit background at RHIC for $1<M<3$ GeV, we thus expect the 
experimental detection of a
 jet-plasma contribution to be feasible. For real photons, the inclusive
  yield is largely dominated by the background coming from the decay of neutral mesons~\cite{phe_pho}, 
 in the region where the jet-thermal contribution is expected to be maximal.

While the thermally induced dilepton yield may be 
strongly affected by  initial conditions like temperature and thermalization time,
this is not the case for the jet-plasma contribution. 
This is due to the phase-space
distribution of jets, which  has a weak dependence on temperature and
makes the dilepton production less pronounced in the early time of the
QGP evolution (see Ref.~\cite{simon2} for a  more detailed discussion 
of the effect of initial conditions on real photon production).  The same argument 
can be also used to estimate the effects of a possible chemical  
non-equilibrium on the jet-plasma contribution.  
On one hand, lower quark fugacities suppress the dilepton
emission at a given temperature, but on the other hand, smaller fugacities
would imply larger temperature, thus increasing the dilepton yield.  
The interplay of those effects was studied in Ref.~\cite{rapp_2001} for 
thermally induced dileptons, showing a suppression in the yield for large invariant masses at RHIC (a factor $\sim$ 2 of suppression for $M\sim$ 4 GeV).
However, for jet-medium interactions the fugacities would
 enter into the production rate only linearly rather than quadratically for
 thermally induced dileptons, thus the effect of chemical non-equilibrium 
 should be less pronounced.

The effect of the momentum cut on single lepton rapidity is shown in Fig.~\ref{y_cut} 
at RHIC energy for Drell-Yan and thermal-thermal processes without HTL 
corrections.  For both cases, the cut reduces the yield by a factor $\sim$ 3 
and is almost independent of $M$, except in the low mass region.  
When $M$ is small, the lepton rapidities tend to be very close to the 
pair rapidity $y_d=0$, making the cut less important.

The effect of energy loss on the jet-thermal lepton pair production 
is explored in Fig.~\ref{htl_jet_eloss} for RHIC.
We observe that for the case without HTL effects (Born term), 
dileptons are reduced by about $30\%$ for $M\sim$ 1 GeV, while the suppression
decreases with increasing invariant mass, reaching about 15\% above $M$ = 4 GeV. For a given invariant mass $M$ and jet energy $E_1$, the minimum energy for 
the thermal parton is $E_{2_{min}}=M^2/4E_1$. This minimal value is then 
favored by the steep thermal spectrum, leading to a dependence 
of the yield that is roughly given by $\exp(-M^2/4E_1T)$.  This implies that dileptons with large 
mass $M$ are more likely to be emitted at times where the temperatures $T$ 
is still high which favors small jet propagation time and 
small energy loss.  Including HTL effects leads 
to a suppression of about $30\%$ for the range dominated by the cut-pole 
contribution, $M<$ 2 GeV and $M>7$ GeV. The region $2<M<7$ GeV, dominated by 
the pole-pole contribution, shows a weaker suppression, equivalent in 
magnitude to that of the Born term in the corresponding invariant mass range. 
The suppression factor is weaker for dileptons 
from jet-medium interactions than for high-$p_T$ pions, where a large
suppression factor ($\sim 4-5$) has been observed at RHIC~\cite{phenix2}.
This can be understood because dileptons can be produced at any point during 
 the propagation of the jet in the medium, whereas pions, due to their large 
 formation time, are produced after a jet
 has left the medium, and thus they suffer from the full loss of energy

\begin{figure}[ht!]
\epsfig{file=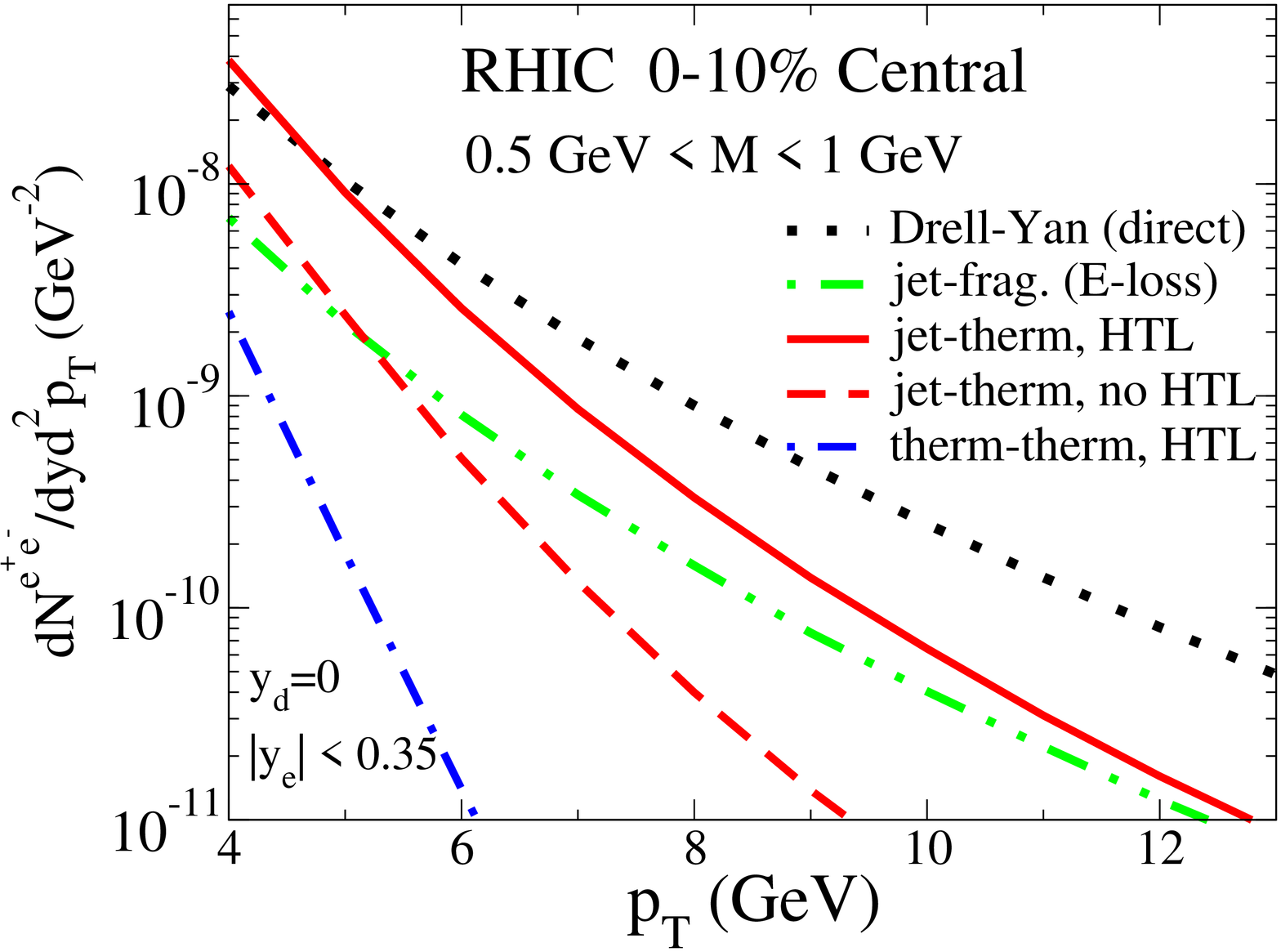,width=6.2cm,angle=-90}
\epsfig{file=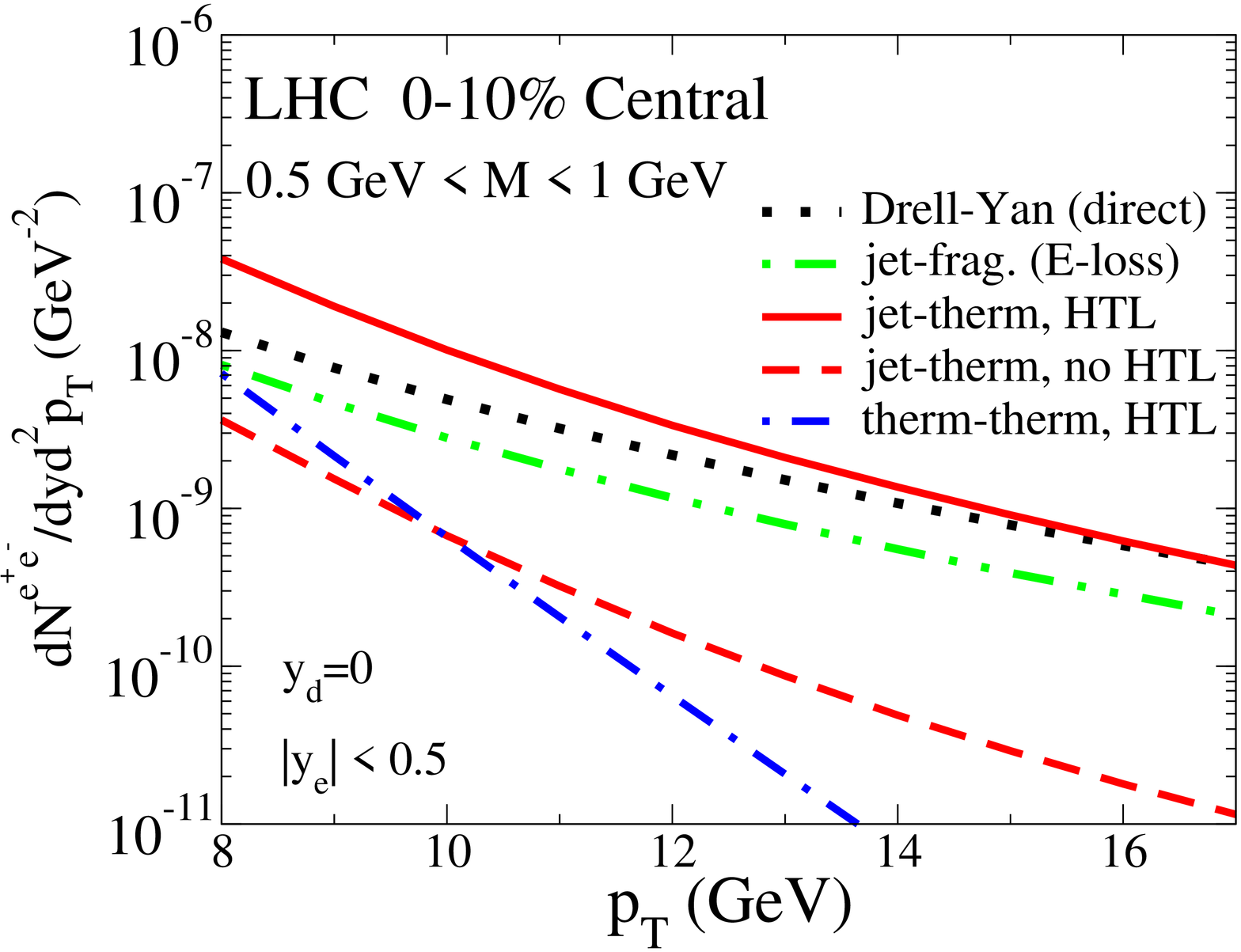,width=6.2cm,angle=-90}
\caption{(Color online) $p_T$ distribution of dileptons, integrated in the range $0.5 < M < 1$ GeV, for RHIC (left) and the LHC (right).  Dotted lines: direct Drell-Yan process; double dot-dashed lines: jet-fragmentation process; solid lines: jet-thermal process with HTL effects; dashed lines: jet-thermal process without HTL effects; and dot-dashed lines: thermal induced reactions with HTL effects.}
\label{dn_dydpt}
\end{figure}

\begin{figure}
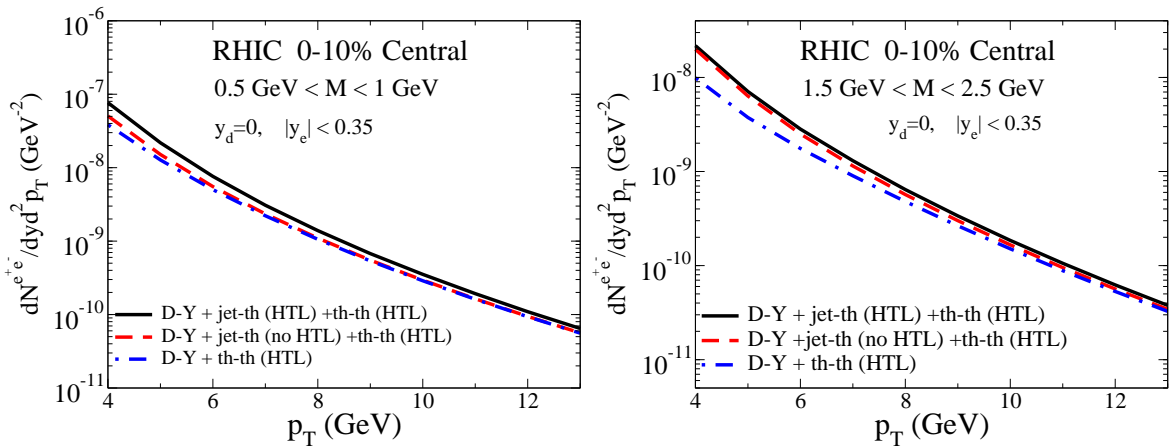

\epsfig{file=dilep_rhic_sum_K.eps,width=7.7cm}
\epsfig{file=dilep_dptsum_rhic_M15_25_K.eps,width=7.7cm}
\caption{(Color online) $p_T$ distribution of dileptons, integrated in the range $0.5 < M < 1$ GeV (left panel) and $1.5 < M < 2.5$ GeV (right panel), for central Au+Au at RHIC.  Solid line: sum of Drell-Yan process, jet-thermal and thermal-thermal process with HTL effects; dashed line: sum of Drell-Yan process, jet-thermal without HTL and thermal-thermal with HTL effects; and dot-dashed line: sum of Drell-Yan process and thermal-thermal reactions with HTL effects. }
\label{sum}
\end{figure}

It is also interesting to discuss the dilepton yield as a function of the 
dilepton transverse momentum $p_T$ in certain windows of the mass $M$.
This is done by substituting the integral over $p_T$ in 
Eqs.~(\ref{yield_final}) and (\ref{D-Y}), by $\int dM^2/(2\pi p_T)$.  
The results for RHIC and the LHC are displayed in Fig.~\ref{dn_dydpt}, for 
the mass integrated in the range $0.5 < M <$1 GeV.  The ordering of the 
contributing sources here is very similar to the one seen for real photons 
as a function of $p_T$ in Ref.~\cite{simon2}. The direct 
component of the Drell-Yan process dominates for high-$p_T$ dileptons at RHIC 
while the jet-thermal contribution, with HTL, dominates for $p_T < $5 GeV. 
The hard thermal loop calculation enhances this yield by more than a factor 
4 compared to the jet-thermal contribution without HTL. At the LHC, 
jet-thermal dileptons (HTL effects included) are the most important source 
in the entire $p_T$ range, $8<p_T<17$ GeV. The jet-thermal interaction 
appears to be as important for dileptons as it was for real photon production.

The total direct dilepton spectrum for RHIC is shown in the left panel of 
Fig.~\ref{sum}. The solid line includes Drell-Yan and QGP contribution 
(jet-thermal and thermal-thermal) with HTL effects.  Leaving out 
the HTL resummation for jet-thermal dileptons (dashed line)
reduces the yield by a factor $\sim 1.5$ around $p_T$=4 GeV. The absence of
any jet-thermal interactions at all (dot-dashed line) would reduce the total 
yield by a factor $\sim 2$ at $p_T$=4 GeV. This emphasizes the importance of 
this process in the presence of a QGP. 

The only potentially important contribution that is not included in our work 
is in-medium bremsstrahlung ($q\,i\rightarrow q\,i\gamma^*$) and 
annihilation ($q\bar{q}i\rightarrow \gamma^*i$) of an incoming thermal parton 
or jet, where $i$ denotes a quark, antiquark or gluon. This goes beyond the current formulation of AMY. However, they have been calculated in 
Ref.~\cite{Aurenche:2002wq} for the case of incoming  {\it thermal} partons,
  showing that for low mass dileptons, the 
bremsstrahlung and annihilation more than double the contribution obtained 
from  $2\rightarrow 2$ 
processes ($q+g\rightarrow q+\gamma^*,q+\bar{q}\rightarrow g+\gamma^*$).  
It thus turns out that thermally induced bremsstrahlung and annihilation 
are as important relatively to $2\rightarrow 2$ processes, no matter if the photons are virtual or real~\cite{Arnold:2001ms}. On the other hand, for the case of real photons and incoming jets, 
it has been shown in Ref.~\cite{simon2} that those bremsstrahlung and 
annihilation processes are reduced by a factor 3-4, relatively to the 
in-medium jet-photon conversion process.  This is because the momentum 
distribution of jets is less steep than the thermal one, making the 
jet-photon conversion the most important in-medium process. Therefore, if one assumes that the 
in-medium jet-bremsstrahlung contributes in the same way to virtual and to
real photons, it would enhance the solid line by less than 15$\%$.

Finally, the right panel of Fig.~\ref{sum} shows the dilepton spectrum for 
an another invariant mass window. It is located at higher values 
$1.5 < M <2.5$ GeV between the $\phi$ and the $J/\psi$ masses.  As we could 
have expected from Fig.~\ref{htl_jet}, the effect of the HTL resummation 
is not very important for this mass window.  However interactions
of jets with the plasma are still a very important source of dileptons, and 
this should be detectable.

\section{Summary and Conclusions}
\label{sec:vii}

In this work we presented calculations of different sources
of lepton pairs in high energy nuclear collisions. We take into account 
Drell-Yan, fragmentation from jets, QGP contributions and 
heavy quark decay.  Hard thermal loop resummation has been included in the 
calculation of the leading order photon self-energy. We explicitly checked
that the imaginary part of this self-energy, evaluated within 
finite-temperature field theory, and a different approach starting from 
relativistic kinetic theory and using the corresponding Feynman amplitudes,
lead to the same results. 
We obtain the jet-plasma interaction by substituting the phase-space 
distribution of one incoming thermal parton, by the distribution of jets.  
While the HTL corrections are important for thermal-thermal processes at 
low-invariant mass, they are important for both low and high invariant mass 
when the incoming parton is a jet.

For the high momentum window investigated here, $p_T > 4$ GeV,
dilepton emissions due to jet-plasma interactions are
found to be much larger than thermal dilepton emission.
 At low to intermediate dilepton mass, productions from jet-plasma interactions are 
comparable in size to the Drell-Yan contribution and constitute a good 
signature for the presence of a quark gluon plasma provided the dominant 
background of heavy quark decay could be subtracted. The AMY formalism has 
been used to account for energy loss of jets in the QCD plasma; this energy degradation reduces the effect of jet-thermal processes by $\sim$ 30\%. 
Further study involving heavy quark energy loss will be needed to obtain 
a better estimate of this channel, together with an explicit calculation of dileptons from medium-induced bremsstrahlung.

\begin{acknowledgments}
We are grateful to G.D Moore for his help with the details of the AMY formalism and for a critical reading of this manuscript. We thank also B. Cole, S.\ Jeon, J.\ S.\ Gagnon, F.\ Fillion-Gourdeau, 
G.D Moore and R.\ Rapp for helpful discussions. This work was supported 
in parts by the Natural Sciences and Engineering Research Council of Canada, 
le Fonds Nature et Technologies du Qu\'ebec and by DOE grant 
DE-FG02-87ER40328.   
\end{acknowledgments}



\begin{thebibliography}{99}
\bibitem{note1}
A numerical error as been found in the program used to perform the calculations in Ref.~\cite{prcdilep}.  The present paper thus constitutes an erratum and an update.
\end{thebibliography}
\end{document}